\documentclass[traditabstract]{aa}

\usepackage{graphicx}
\usepackage[varg]{txfonts}
\usepackage{hyperref}
\usepackage{natbib,twoopt}
\usepackage{lscape}
\usepackage{subcaption}
\usepackage{nicefrac}
\bibpunct{(}{)}{;}{a}{}{,} 

\makeatother\bibpunct[, ]{(}{)}{;}{a}{}{,}

\newcommand{\farc}{\hbox{$.\!\!^{\prime\prime}$}} 
 
\newcommand{\erg}{$\rm{erg~cm^{-2}~s^{-1}}$} 

\newcommand{\hb}{H$\beta$} 
\newcommand{\ha}{H$\alpha$}
\newcommand{\hg}{H$\gamma$} 
\newcommand{\hd}{H$\delta$} 
 
\newcommand{\hii}{\mbox{H~{\sc ii}}} 
 
\newcommand{\oh}{12+\log(\mathrm{O/H})}

\newcommand{\hei}{\ion{He}{i}} 
\newcommand{\oi}{[\ion{O}{i}]} 
\newcommand{\sii}{[\ion{S}{ii}]} 
\newcommand{\siii}{[\ion{S}{iii}]} 
\newcommand{\oii}{[\ion{O}{ii}]}
\newcommand{\cii}{[\ion{C}{ii}]} 
\newcommand{\oiii}{[\ion{O}{iii}]}
\newcommand{\neiii}{[\ion{Ne}{iii}]}
\newcommand{\nii}{[\ion{N}{ii}]}

\newcommand{\Msun}{$M_\odot~$}
\newcommand{\Msunyr}{$M_\odot~\rm{yr}^{-1}$}

\begin{document}

\title{Hot gas around SN~1998bw: Inferring the progenitor from its environment\thanks{Based on observations collected at the ESO Paranal observatory under ESO program 095.D-0172(A) and data obtained from the ESO Science Archive Facility.}}
\titlerunning{Hot gas around SN98bw}

\author{T.~Kr\"{u}hler\inst{1}
\and H.~Kuncarayakti\inst{2, 3, 4, 5}
\and P.~Schady \inst{1} 
\and J.~P.~Anderson \inst{6}
\and L.~Galbany \inst{7}
\and J.~Gensior \inst{8}}

\institute{Max-Planck-Institut f\"{u}r extraterrestrische Physik, Giessenbachstra\ss e, 85748 Garching, Germany
\and Finnish Centre for Astronomy with ESO (FINCA), University of Turku, V\"ais\"al\"antie 20, 21500 Piikki\"o, Finland
\and Tuorla Observatory, Department of Physics and Astronomy, University of Turku, V\"ais\"al\"antie 20, 21500 Piikki\"o, Finland
\and Millennium Institute of Astrophysics, Casilla 36-D, Santiago, Chile
\and Departamento de Astronom\'ia, Universidad de Chile, Casilla 36-D, Santiago, Chile
\and European Southern Observatory, Alonso de C\'{o}rdova 3107, Vitacura, Casilla 19001, Santiago 19, Chile 
\and PITT PACC, Department of Physics and Astronomy, University of Pittsburgh, Pittsburgh, PA 15260, USA
\and School of Physics and Astronomy, University of Edinburgh, Peter Guthrie Tait Road, Edinburgh EH9 3FD, UK
}

\abstract{Spatially resolved spectroscopy of the environments of explosive transients carries detailed information about the physical properties of the stellar population that gave rise to the explosion, and thus the progenitor itself. Here, we present new observations of ESO184-G82, the galaxy hosting the archetype of the $\gamma$-ray burst/supernova connection, GRB~980425/SN~1998bw, obtained with the integral field spectrograph MUSE mounted at the Very Large Telescope. These observations have yielded detailed maps of emission-line strength for various nebular lines along with physical parameters such as dust extinction, stellar age, and oxygen abundance on spatial scales of 160~pc. The immediate environment of GRB~980425 is young (5~--~8~Myr) and consistent with a mildly extinguished ($A_V\sim0.1~\mathrm{mag}$) progenitor of zero-age main-sequence mass between 25~\Msun and 40~\Msun and an oxygen abundance {$\oh\sim8.2$} ($Z\sim0.3~Z_\odot$), which is slightly lower than that of an integrated measurement of the whole galaxy ($\oh\sim8.3$) and a prominent nearby \hii\ region ($\oh\sim8.4$). {This region is significantly younger than the explosion site, and we argue that a scenario in which the GRB progenitor formed in this environment and was subsequently ejected appears very unlikely}. We show that empirical strong-line methods based on \oiii~and/or \nii~ are inadequate to produce accurate maps of oxygen abundance at the level of detail of our MUSE observation as these methods strongly depend on the ionization state of the gas. The metallicity gradient in ESO184-G82 is $-0.06$~dex~kpc$^{-1}$, indicating that the typical offsets of at most few kpc for cosmological GRBs on average have a small impact on oxygen abundance measurements at higher redshift.}

\maketitle

\section{Introduction}
\label{sec:Intro}

Line emission from recombination of ionized hydrogen or from the decay of collisionally excited states of metal ions {is a fundamental tracer of the physical conditions in \hii\ regions}. The absolute and relative intensities of these transitions crucially depend on the ionizing source, electron density in the plasma, ionization state of the elements, and gas-phase abundances \citep{1989agna.book.....O}. This makes emission-line spectra of astronomical sources one of the most elementary diagnostics of galaxy formation and evolution \citep[e.g.,][]{2004ApJ...613..898T, 2006ApJ...644..813E, 2009ApJ...706.1364F}. The total intensity of the hydrogen recombination lines, for example, is proportional the number of O-type stars and thus traces the star formation rate at timescales of $\sim10$~Myr \citep[e.g.,][]{1998ARA&A..36..189K}. The continuum emission at the wavelength of \ha~ in turn originates from B- or A-type stars, which makes the \ha~ equivalent width (EW) a good proxy for the age of the stellar population \citep{1999ApJS..123....3L, 2013ApJ...779..170L}.

Metal abundances have been measured through ratios of prominent emission lines from ions such as O$^{+}$, O$^{2+}$, N$^{+}$, S$^{+}$, and/or recombination lines of hydrogen \citep{1979MNRAS.189...95P, 1979A&A....78..200A}. Given their fundamental importance in galaxy evolution and cosmology, these abundance determinations through nebular emission lines have been the focus of a large body of literature \citep[e.g.,][]{2004ApJ...617..240K, 2005ApJ...631..231P, 2006A&A...454L.127S, 2006A&A...448..955I, 2008ApJ...681.1183K}.

It is thus immediately clear that an emission-line spectrum of cosmological sources carries detailed information about the underlying stellar population and thus has been used to infer properties not only of galaxies but also of the progenitors of explosive transients. Global \citep[e.g.,][]{2008ApJ...673..999P, 2011MNRAS.412.1441L} or local \citep[e.g.,][]{2010MNRAS.407.2660A, 2011ApJ...731L...4M, 2011A&A...530A..95L} properties of nearby supernova (SN) hosts and cosmological $\gamma$-ray bursts \citep[GRBs; e.g.,][]{2007A&A...464..529W, 2012A&A...546A...8K, 2013ApJ...774..119G} or superluminous supernovae \citep[e.g.,][]{2013ApJ...763L..28C, 2014ApJ...787..138L, 2014arXiv1409.8331L, 2016arXiv160408207P} have likewise been used to compare progenitor models with the expected environments. 

A fundamental assumption of all these studies is the hypothesis that there is a tight relation between the two primary observables, gas-phase oxygen abundance and age of \hii~regions, and the progenitor properties of metallicity and lifetime and thus initial mass. Clearly, this link is most robust when coming from an analysis of the co-spatial stellar population. Integral field spectroscopy (IFS) with high angular resolution is thus arguably the most comprehensive way of studying the environments of explosive transients. Low-redshift galaxies hosting supernovae (SNe), for example, are hence ideal targets for state-of-the-art integral field units \citep[IFUs; e.g.,][]{2013AJ....146...30K, 2013AJ....146...31K, 2014A&A...572A..38G}.

In contrast to SNe, the vast cosmological distances of GRBs \citep[e.g.,][]{2009ApJS..185..526F, 2009Natur.461.1254T, 2012ApJ...758...46K} have always posed serious limitations on using IFS for GRB-selected galaxies. Only a few GRBs are close enough such that the spatial resolution achievable with modern ground-based instrumentation yields constraints on spatial scales better than a kpc. Spatially resolved spectroscopy has therefore only been obtained for a handful of nearby GRB hosts using long-slit spectroscopy \citep[e.g.,][]{2008ApJ...676.1151T, 2011ApJ...739...23L, 2015A&A...579A.126S} or the previous generation of IFUs \citep{2008A&A...490...45C, 2014MNRAS.441.2034T}.

\begin{figure}
\begin{subfigure}{.48\textwidth}
  \includegraphics[width=0.999\linewidth]{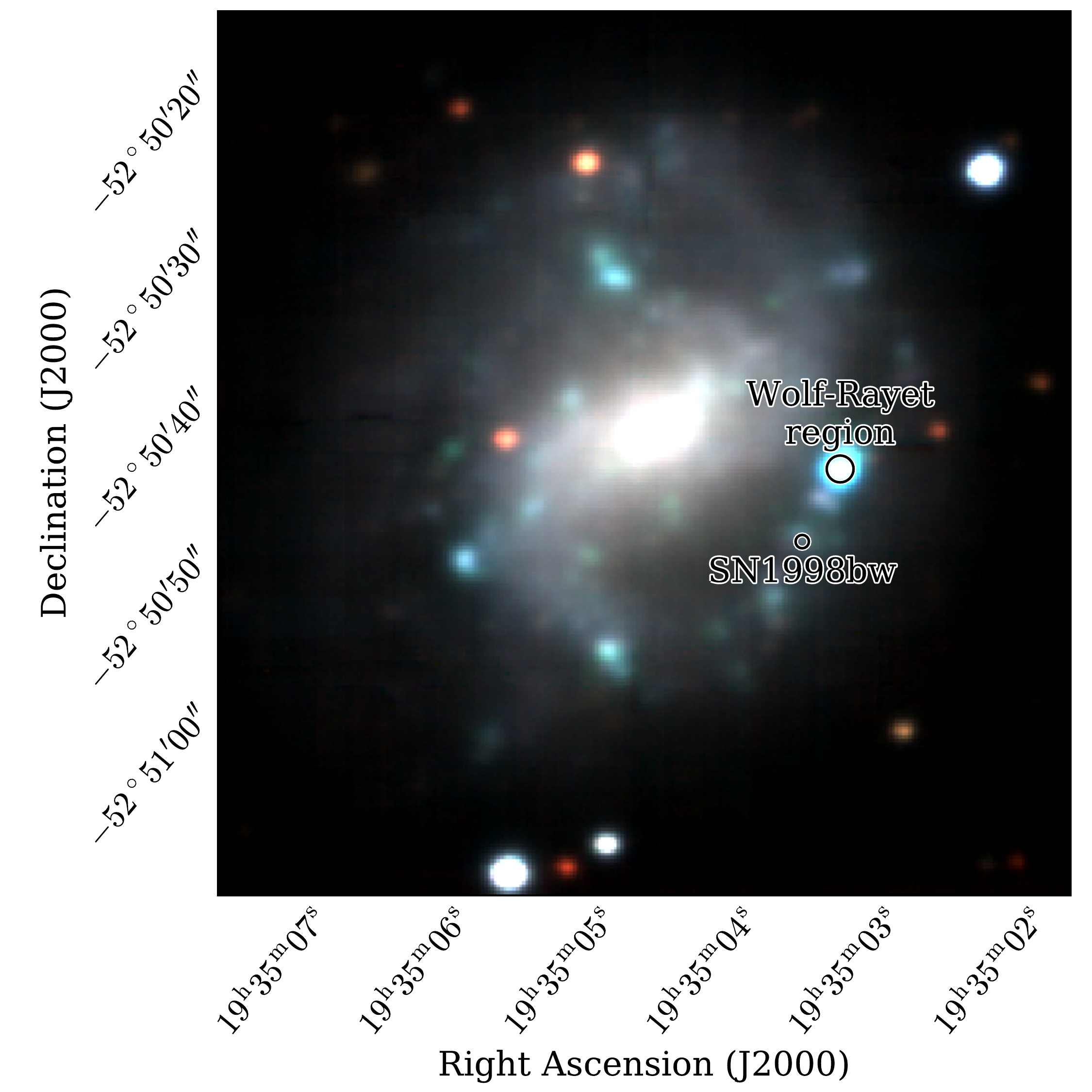}
\end{subfigure}
\caption{False-color composite from reconstructed $VRI$-band images from the MUSE data cube. The image spans approximately 55\arcsec~by 55\arcsec, or 10 by 10~kpc at $z=0.0086$.}
\label{fig:Host}
\end{figure}

The focus of this article is on our new observations of the poster child of the connection between GRBs and SNe, GRB~980425/SN~1998bw, using the Multi-Unit Spectroscopic Explorer (MUSE; \citealp{2010SPIE.7735E..08B}). GRB~980425 is the closest GRB yet discovered; the GRB \citep[e.g.,][]{1998Natur.395..670G, 1998Natur.395..663K}, SN \citep[e.g.,][]{1998Natur.395..672I, 2001ApJ...555..900P, 2006ApJ...640..854M}, and host galaxy \citep[e.g.,][]{2000ApJ...542L..89F, 2005NewA...11..103S, 2006A&A...454..103H, 2009ApJ...693..347M, 2014A&A...562A..70M, 2016arXiv160901742M, 2012ApJ...746....7L, 2015MNRAS.454L..51A} are extensively discussed in existing literature. 

Compared to the bulk of cosmological GRBs, GRB~980425 is rather peculiar: the isotropic-equivalent release in $\gamma$-rays of GRB~980425 was $\sim10^{48}$~erg \citep{1998Natur.395..670G}, which is a factor of ten lower than other local, low-luminosity GRBs, or around five orders of magnitudes less than conventional, higher redshift GRBs \citep{2013ApJ...776...98X}. No bright multiwavelength afterglow was observed for GRB~980425 despite its proximity. However, the associated SN without hydrogen or helium in its spectrum, broad metal absorption lines, and high luminosity has proven to be typical of GRB-related SNe in general \citep{2012grbu.book..169H}.

The host galaxy of GRB~980425/SN~1998bw, ESO184-G82 \citep{1989spce.book.....L}, is a barred spiral dwarf galaxy \citep{2000ApJ...542L..89F} seen nearly face on and shown in Fig.~\ref{fig:Host}. It has a visible {extent} of approximately 67\arcsec~by 57\arcsec~(12 x 10~kpc) at the $B=26.5$~mag isophote \citep{2005NewA...11..103S}. Its brightness, luminosity, and stellar mass are $B=14.94$~mag, $M_B=-17.65$~mag or $L=0.05~{L}^{\star}$, and $\log (M_{*}/M_{\odot})= 8.7 $, respectively \citep{2005NewA...11..103S, 2014A&A...562A..70M}. SN~1998bw exploded in an \hii~region 12\arcsec\ distant (2~kpc projected) from its center and 860~pc to the southeast of a young star-forming region that displays signatures of Wolf-Rayet (WR) stars in its spectrum \citep{2006A&A...454..103H}, the so-called Wolf-Rayet region (Fig.~\ref{fig:Host}).

Despite the large set of recent literature on GRB~980425, SN~1998bw and its host mentioned above, we summarize our new data and conclusions here for three main reasons: First, the unique combination of spatial resolution and sensitivity of MUSE yields detailed maps of emission-line strength, dust reddening, and oxygen abundance, which help us to clarify some of the ambiguities around SN~1998bw and its host from previous works. Second, these maps provide the tightest and most accurate constraints on the immediate environment and underlying stellar population of SN~1998bw yet available and thus allow us to infer the progenitor properties of the GRB from its parent stellar population. And last, our new data offer an informative example of spatially resolved oxygen-abundance measurements in star-forming galaxies through strong-line diagnostics and their dependence on other physical conditions in the interstellar medium (ISM).

Throughout the paper, we adopt a flat $\Lambda$CDM cosmology with Planck parameters ($H_0=67.3~\mathrm{km}~\mathrm{s}^{-1}~\mathrm{Mpc}^{-1}$, $\Omega_\mathrm{m}$=0.315, $\Omega_\Lambda$=0.685; \citealt{2014A&A...571A..16P}), a \citet{2003PASP..115..763C} initial mass function (IMF), solar abundances from \citet{2009ARA&A..47..481A}, and report errors at the $1\sigma$ confidence level.

\section{Observations}

We observed ESO184-G82 ($z=0.0086$, or $D_L=37$~Mpc) using the Multi-Unit Spectroscopic Explorer (MUSE; \citealp{2010SPIE.7735E..08B}) at the ESO Very Large Telescope (VLT) during the two clear nights starting on 2015-05-14 and 2015-05-15 in a classical observing run from Paranal. On each night, we obtained four dithered exposures of 450~s integration each, totaling 3600~s on source. The on-target frames were supplemented by an offset pointing to blank sky for 200~s. For absolute flux calibration, the spectrophotometric standard LTT3218 was observed at the beginning of each night. The full width half maximum of the stellar point spread function, which defines our spatial resolution, is between 0\farc{9} (at 9000~\AA) and 1\farc{1} (at 5000~\AA) in the MUSE data.

The MUSE instrument is a state-of-the-art integral field spectrograph that splits the light into 24 identical subunits. In the wide-field mode, each of these sub-IFUs disperses a $60^{\prime\prime}\times 2.5^{\prime\prime}$ patch of the sky onto a single CCD. In this way, MUSE covers a continuous sky region of $60^{\prime\prime}\times 60^{\prime\prime}$ in the wavelength range between 4750~\AA~and 9300~\AA~when operated in its nominal configuration. With its excellent total throughput, small spaxel size (0\farc{2}~$\times$~0\farc{2}), and decent resolving power ($1800 < R < 3600$ increasing from blue to red wavelengths), MUSE offers an unprecedented combination of sensitivity, spatial resolution, and field of view for IFUs \citep{2010SPIE.7735E..08B}.

\section{Data reduction}
\label{sec:red}
We reduced the MUSE data with the pipeline supplied through ESO\footnote{http://www.eso.org/sci/software/pipelines/} in its version \texttt{1.2.1} \citep{2014ASPC..485..451W}, which applies corrections for bias level, flat-fields, illumination level, and geometric distortions. The pipeline also performs the wavelength calibration using daytime arc-lamp frames, which is subsequently refined by skylines in the science data. The sky background was subtracted using an offset pointing to blank sky and making use of algorithms from the Zurich Atmospheric Package \citep{2016MNRAS.458.3210S}. The exposures from the two different nights were then corrected for slight pointing offsets between night one and two, stacked using variance weighting, and dereddened based on the Galactic foreground $E_{B-V}=0.05$~mag \citep{2011ApJ...737..103S} assuming an average Milky Way extinction law \citep{1992ApJ...395..130P} and $R_V=3.08$. 

{The spectrum of the star at $\mathrm{RA(J2000)=19^{h}35^{m}02^{s}.00}$, $\mathrm{Decl(J2000)} = -52\degr50\arcmin21\farc{1}$ was used to correct for telluric absorption via \texttt{molecfit} \citep{2015A&A...576A..77S}. By fitting the three prominent telluric absorption bands within the MUSE wavelength coverage (centered around 6870~\AA, 7600~\AA, and 7630~\AA) with a physical model of the atmospheric molecular oxygen and water vapor content, we derived the telluric absorption for this single star, which we then subsequently applied to all spaxels.}

The final data cube has slight astrometric offsets, which we corrected by tying the position of stars in the field of MUSE to coordinates from a reference image taken on 2000-10-25 with the SOFI imager on the ESO New Technology Telescope. We then measured the position of the SN in the reference frame, mapping it onto the MUSE cube with an accuracy of around 50~mas. Figure~\ref{fig:Host} shows a false-color image reconstructed from the MUSE cube where the position of SN~1998bw derived from the reference image is indicated.

Similarly, we use existing photometry to corroborate our flux calibration through the $V$-, $R_C$- and $I_C$-band magnitudes of star 1 of \citet{2011AJ....141..163C} and synthetic photometry from the MUSE data cube, yielding differences of $\Delta V = 0.05\pm0.03$~mag, $\Delta R_C = 0.05\pm0.06$~mag, and $\Delta I_C = 0.00\pm0.05$~mag. After applying a linear fit in wavelength to these correction terms, we can accurately reproduce the optical colors of the host galaxy \citep{2005NewA...11..103S} to better than 0.02~mag.

\section{Analysis and discussion}

\subsection{Separating gas-phase and stellar component}
\label{sec:stargas}

\begin{figure}
\includegraphics[angle=0, width=0.99\columnwidth]{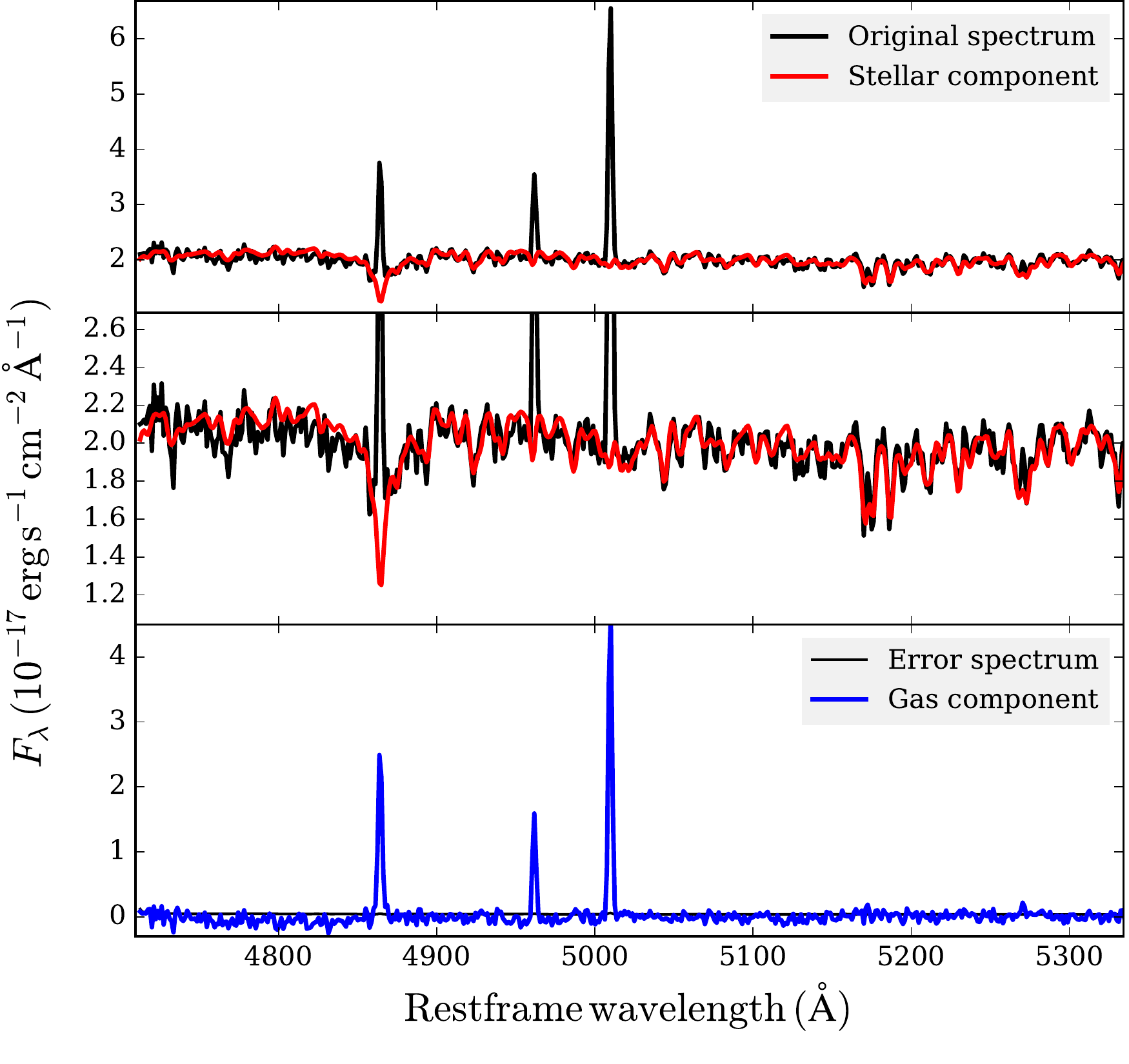}
\caption{Separating stellar and gas-phase components at the {wavelength around \hb~ and \oiii($\lambda\lambda4959,5007$). \textit{Top}: Original spectrum (black) and  fitted stellar component (red). \textit{Middle}: Zoom-in to the continuum. \textit{Bottom}: Resulting spectrum of the gas-phase contribution (blue) together with the error spectrum (grey).}}
\label{fig:stargas}
\end{figure}

As we are primarily interested in the absolute and relative strengths of the nebular lines and thus the ionized gas component of the galaxy, we needed to remove the stellar Balmer absorption for accurate line flux measurements, in particular for the \hb~line (Fig.~\ref{fig:stargas}). The strength of the stellar absorption is primarily a function of the age of the underlying stellar population. It thus depends on the position within a galaxy and needs to be accurately modeled for reliable constraints on the Balmer decrement, which we use to measure dust reddening maps.

We separated the stellar and gas-phase components of the galaxy by fitting a linear superposition of template spectra, based on the \citet{2003MNRAS.344.1000B} models, to the MUSE data. We divided the full field of view into regions with a size of $0\farc{4}\times0\farc{4}$ (or $2 \times 2$ spaxels) and extracted spectra for each of the $\sim22000$ subregions. These spectra were then fit with stellar population models using \texttt{starlight} \citep{2005MNRAS.358..363C, 2009RMxAC..35..127C} following methods that we previously applied to MUSE data \citep{2016MNRAS.455.4087G, 2016arXiv160703446K, 2016arXiv160900013P}. The $2\times2$ co-adding effectively increases the signal-to-noise ratio (S/N) at the expense of spatial resolution for the stellar properties, but is necessary to robustly perform an automated fit in particular in the fainter regions of the galaxy. We then linearly scale the best-fit stellar template to the intensity in single spaxels. Subtracting this stellar component from the original data leaves us with the spectral contribution of the gas-phase only (Fig.~\ref{fig:stargas}).

After having isolated the stellar component from the observed spectrum, it is trivial to produce maps of line flux (integral over the nebular emission line), continuum (average of the stellar component), and equivalent width (flux divided by continuum). 

\begin{figure}
\begin{center}
  \includegraphics[width=0.99\linewidth]{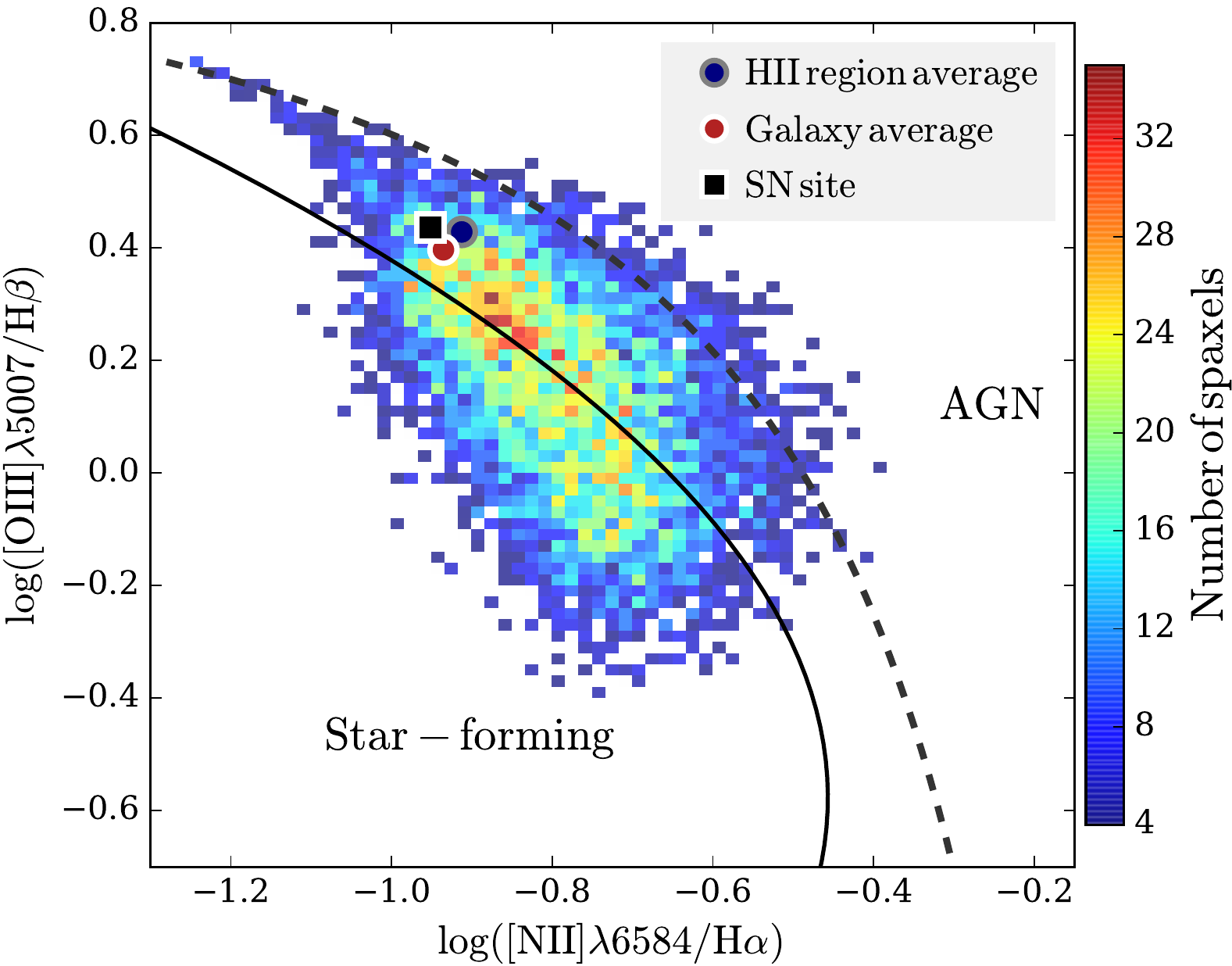}
\caption{Spaxel BPT diagram for ESO184-G82. The solid line denotes the ridge line of SDSS galaxies \citep{2008MNRAS.385..769B}, whereas the dotted line represents the $z=0$ classification line between star formation and AGN ionization \citep{2013ApJ...774..100K}. The  explosion site, a galaxy average (including all spaxels), and an \hii-region average (including only spaxels with EW$_{\mathrm{H\alpha}}>10$~\AA) are indicated.}
\label{fig:BPT}
\end{center}
\end{figure}

Plotting the characteristic emission-line flux ratios of \oiii/\hb~ versus \nii/\ha~ for individual spaxels in the BPT diagram \citep{1981PASP...93....5B} allows us to immediately discard active galactic nuclei or shocks as the ionization source and to confirm that the radiation from massive stars is the origin of the nebular lines in ESO184-G82 (Fig.~\ref{fig:BPT}).

\subsection{Equivalent width maps and stellar population ages}

\begin{figure}
\begin{subfigure}{.2425\textwidth}
  \includegraphics[width=1.0\linewidth]{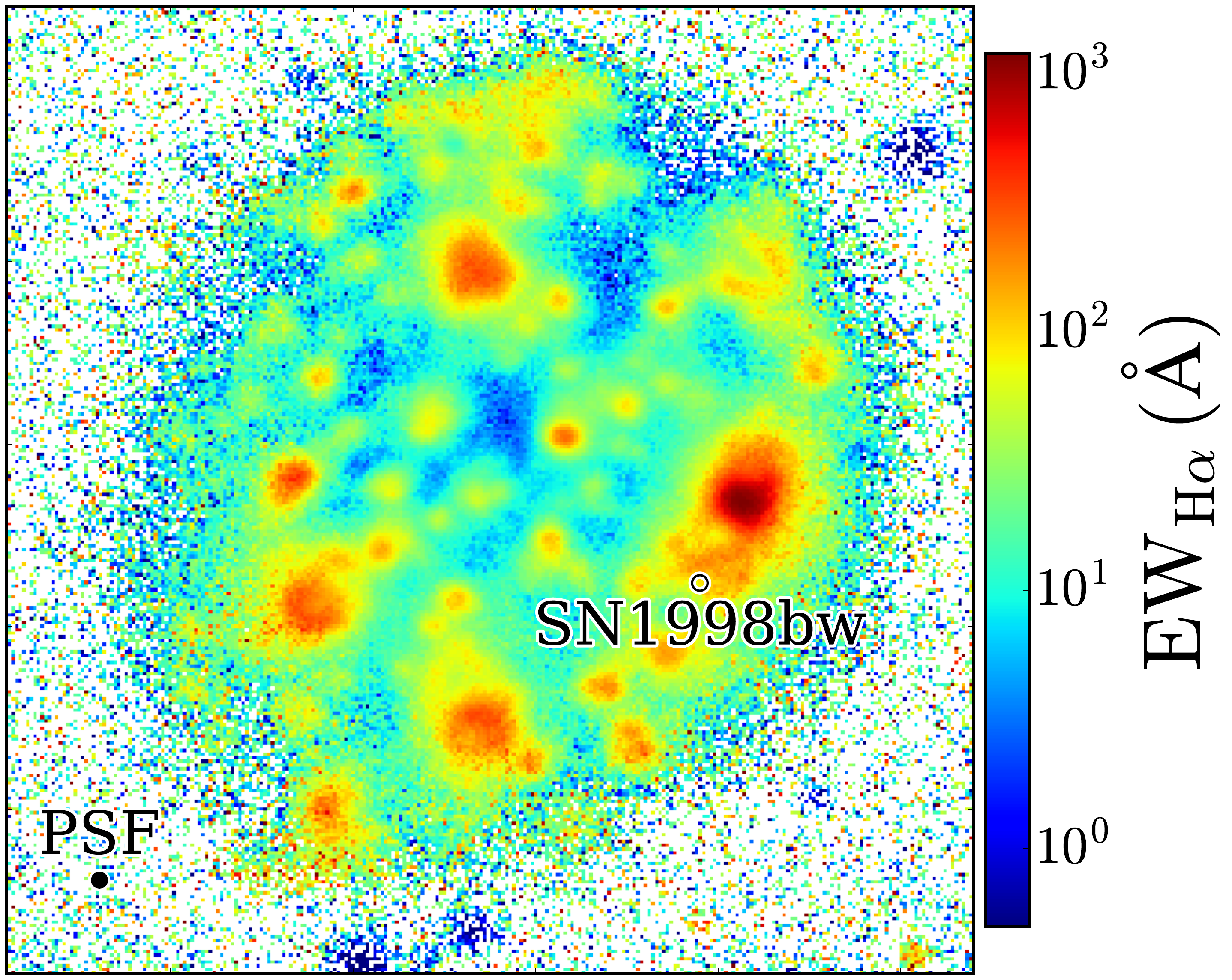}
\end{subfigure}
\begin{subfigure}{.2425\textwidth}
  \includegraphics[width=1.0\linewidth]{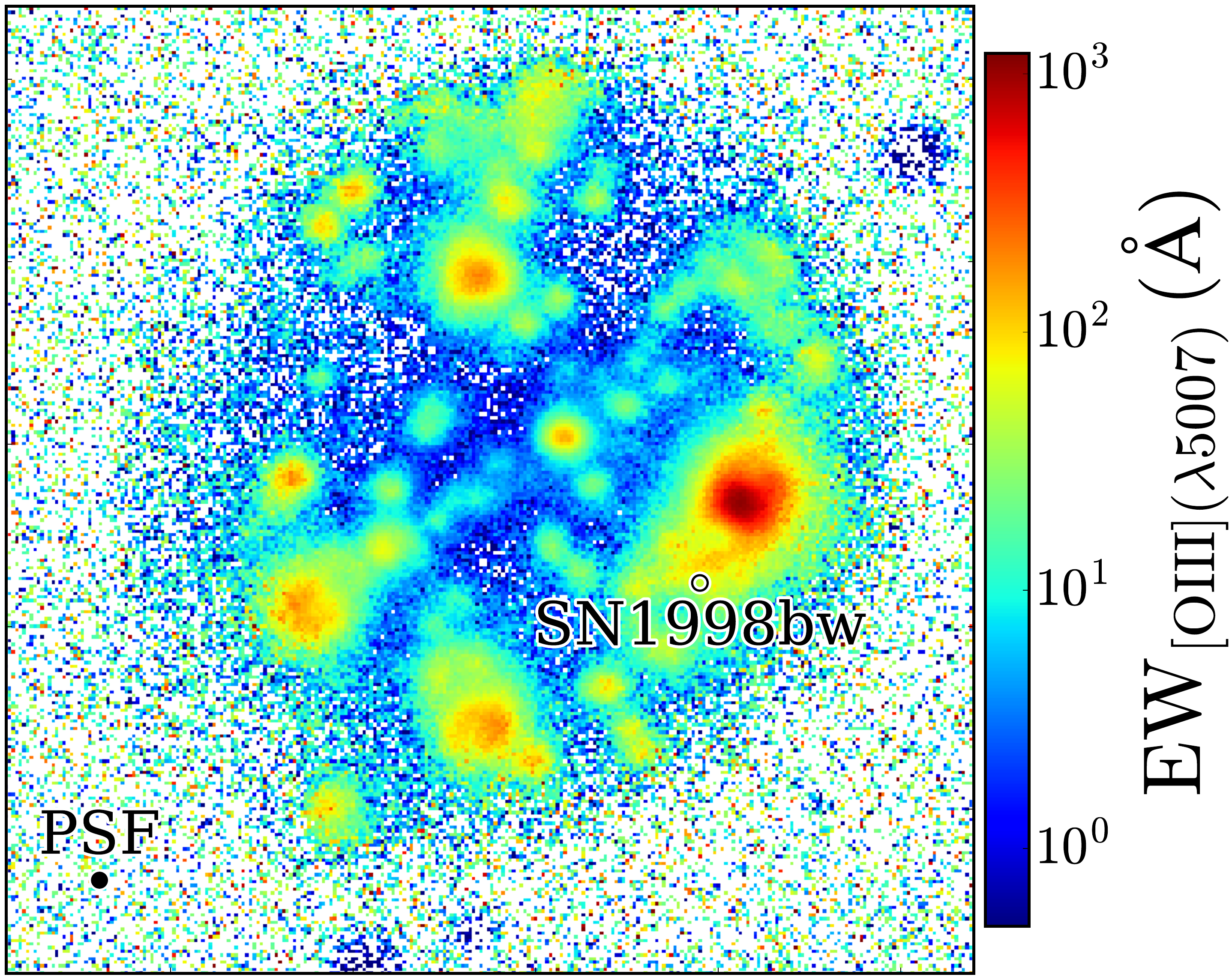}
\end{subfigure}
\begin{subfigure}{.2425\textwidth}
  \includegraphics[width=1.0\linewidth]{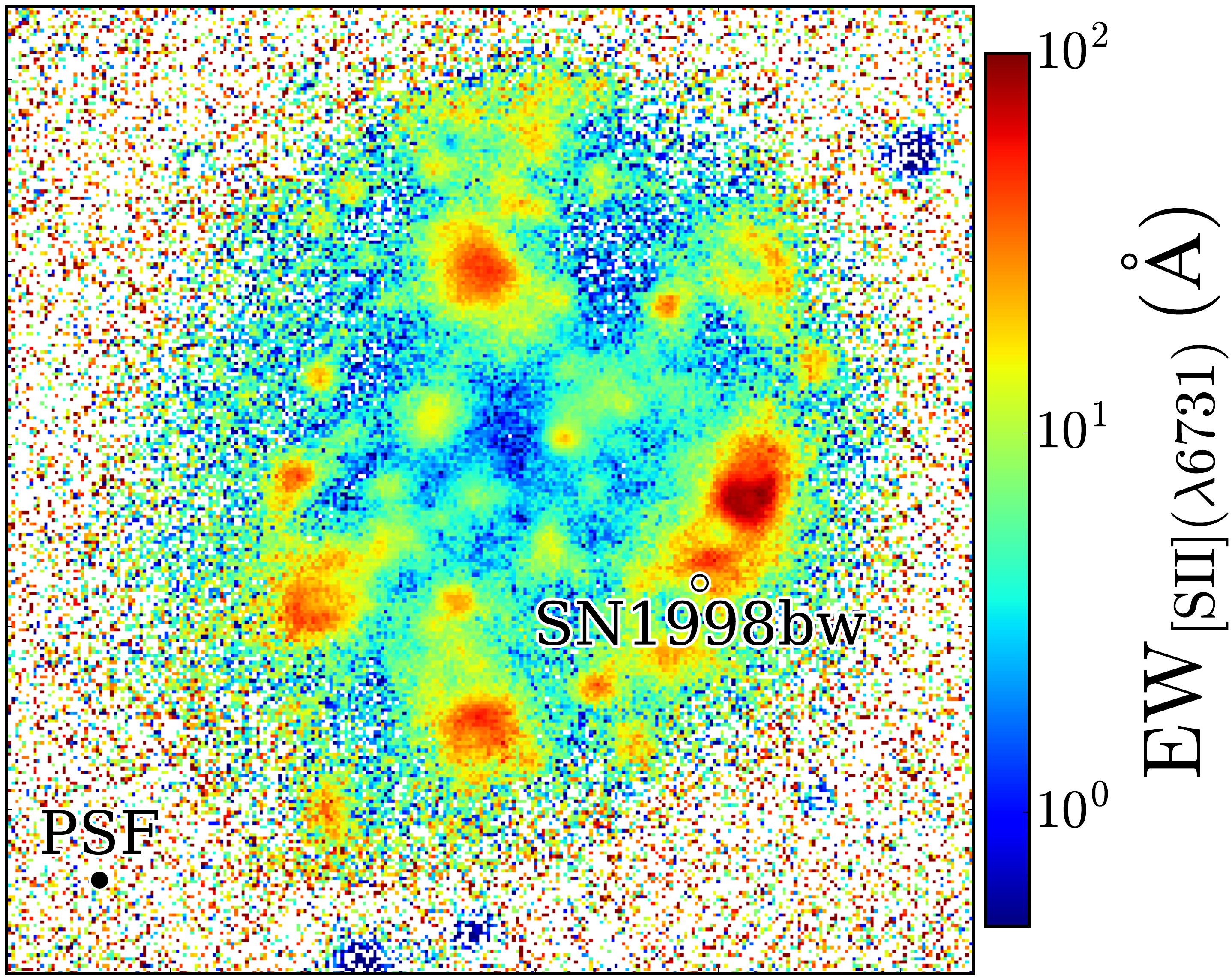}
\end{subfigure}
\begin{subfigure}{.2425\textwidth}
  \includegraphics[width=1.0\linewidth]{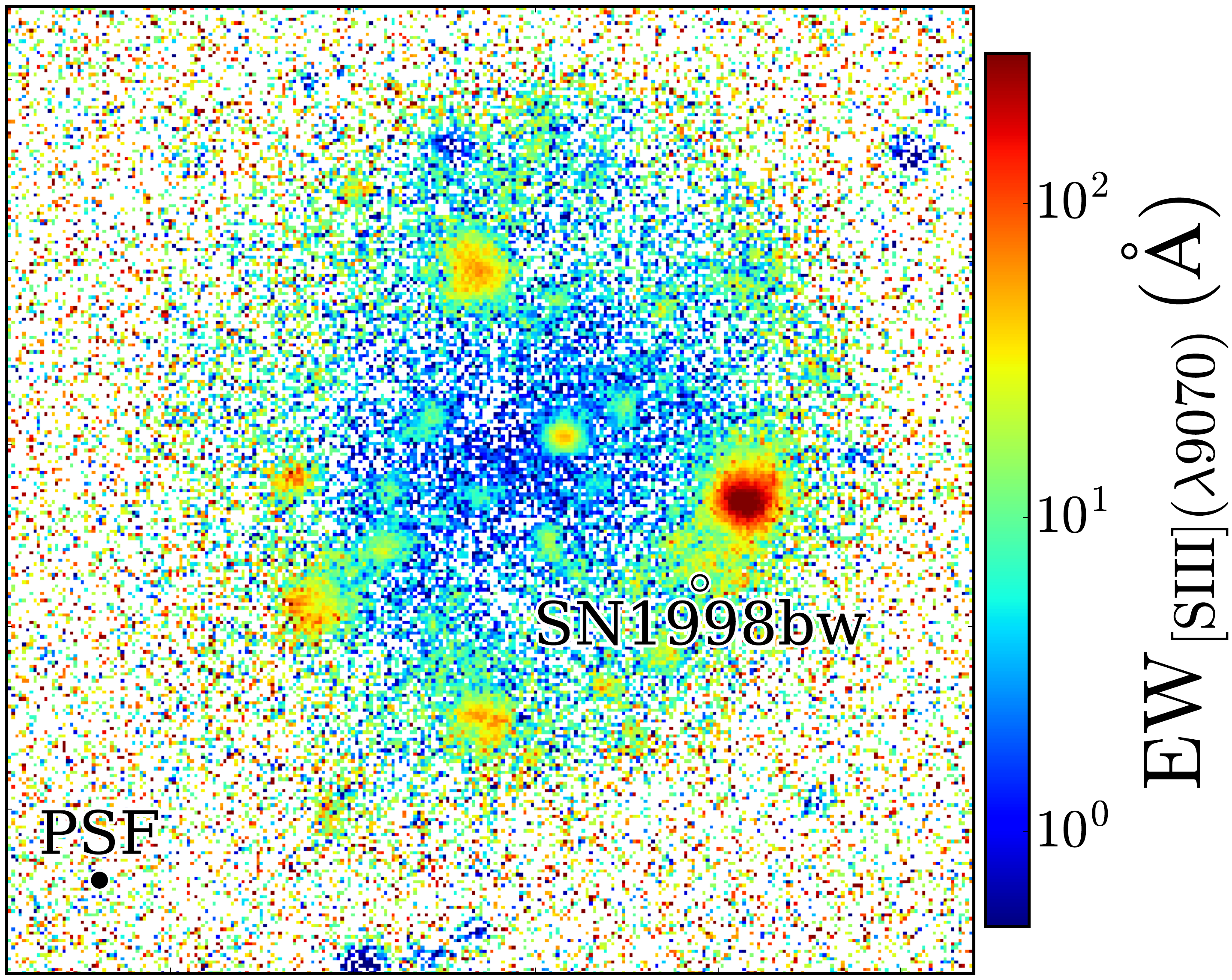}
\end{subfigure}
\caption{Reconstructed images from the MUSE data cube. Each panel shows the host of GRB~980425 in a different nebular line. \textit{Top left}: \ha. \textit{Top right}: \oiii($\lambda$5007). \textit{Bottom left}:  \sii($\lambda$6731). \textit{Bottom right}:  \siii($\lambda$9070). All panels are approximately 55\arcsec~by 55\arcsec, or 10~kpc by 10~kpc at the GRB redshift. The effective spatial resolution is given by the point spread function (PSF) indicated in the lower left corner of each image with a full width half maximum (FWHM) of approximately 0\farc{9} or 160~pc.}
\label{fig:EW}
\end{figure}

The \ha~equivalent width map, which is a rather direct proxy of stellar population age, in particular for young stellar populations, is shown in Fig.~\ref{fig:EW}. The spaxel closest to the SN position has an \ha~equivalent width of EW$_{\mathrm{H\alpha}}=98\pm4$~\AA. The spaxels within a radius of 70~pc yield EW$_{\mathrm{H\alpha}}=92\pm15$~\AA. 

Assuming a single stellar population from an instantaneous starburst, this EW$_{\mathrm{H\alpha}}$ corresponds to stellar population ages between 5 Myr and 8 Myr\footnote{{Constraints from other, age-sensitive lines, such as H$\beta$ or \hei~\citep{1999ApJS..125..489G}, and the stellar continuum (Sect.~\ref{app:fors}) generally agree with this age range.}} from various models at metallicities of $Z=0.2$~$Z_{\odot}$ \citep[see, e.g.,][and references therein]{2013ApJ...779..170L, 2016arXiv160703446K}. The relatively large range in age is almost entirely due to the spread from different stellar evolution models or initial mass functions. These ages corresponds to lifetimes of stars with zero-age main sequence (ZAMS) masses between approximately 25 and 40~\Msun \citep{1994A&AS..105...29F, 2005A&A...429..581M}. This is consistent with the ejected oxygen mass of the SN ($M_{\mathrm{O}}\sim5-6~M_{\odot}$ evolved from a $M_{\rm{ZAMS}} \sim 30-40$~\Msun star) derived through modeling the SN~1998bw nebular spectra \citep{2001ApJ...559.1047M, 2006ApJ...640..854M}.

These considerations are only valid, of course, if the progenitor was born where it exploded, and was not ejected from the nearby WR region \citep{2006A&A...454..103H}. However, this region is so young that timing arguments make an ejection scenario quite contrived: Very high EW values of \ha~and nebular transitions are observed in the center of the WR region (EW$_{\mathrm{H\alpha}}>900~\AA$ and EW$_{\oiii(\lambda5007)}>850~\AA)$\footnote{These are strictly lower limits. The fact that the compact WR region is convolved with the seeing-introduced spatial scale of $\sim$0\farc{9} leads to a smoothed EW distribution. In fact, archival VLT spectroscopy discussed in the Appendix were taken under significantly better atmospheric conditions and yield EW$_{\oiii(\lambda5007)}\sim2000$~\AA~(Appendix~\ref{app:fors}).}. {Together with the detection of strong \hei($\lambda4922$) with an EW$_{\mathrm{HeI(\lambda4922)}}=3.5\pm0.2~\AA$}, they ascertain population ages younger than 3 Myr (see also Appendix~\ref{app:fors}) in instantaneous starburst models, or $M_{\mathrm{ZAMS}} \gtrsim 60$~\Msun \citep[see, e.g.,][and references therein for a similar case]{2015MNRAS.451L..65T}. The discrepancy with respect to the progenitor mass from SN modeling \citep[e.g.,][]{2001ApJ...559.1047M, 2006ApJ...640..854M} then suggests that the GRB progenitor was indeed not born in the WR region, but rather formed in situ.

To travel to the explosion site in less time than the age of the WR region ($<3$~Myr), the progenitor would require very high peculiar velocities $v$ of $v>260~\mathrm{km~s^{-1}}$. This is again a strict lower limit, as projection effects further increase the necessary velocities. Scenarios that are believed to give rise to these kinds of massive runaway stars are dynamical few-body encounters or binary supernovae, but both seem unfeasible here. A dynamical ejection produces hyper-velocity stars in only very rare and extreme cases \citep{2001A&A...365...49H, 2012ApJ...751..133P}, and the probability of potential GRB progenitors ($M_{\rm{ZAMS}}\gtrsim 20~M_\odot$) receiving velocity kicks of $>200~\mathrm{km~s^{-1}}$ from a companion SN is also practically zero \citep{2011MNRAS.414.3501E}. This makes a binary supernova origin highly implausible as there simply would not be enough time to evolve and explode the primary and eject the secondary to a distance $\gtrsim860$~pc. Also the fraction of stars ejected by dynamical encounters is of course a function of the elapsed time after starburst, and reaches only 0.01 or 0.03 at 1~Myr or 3~Myr at $M_{\rm{ZAMS}}\sim 35~M_\odot$ \citep{2012ApJ...746...15B}, again leaving little time for the ejected star to travel as far as 860~pc (or further).

Given the presence of massive stars in the vicinity of the SN position \citep{2000ApJ...542L..89F}, the substantial level of recent star formation, as evidenced by high EWs of nebular lines at the SN position (Fig.~\ref{fig:EW}), and the consistency between $M_{\rm{ZAMS}}$ derived from the age of the stellar population and the SN~1998bw nebular spectra, we see no compelling reason to invoke an artificial ejection from the nearby \hii\  region to explain the GRB location within its host.

\subsection{Dust distribution}

\begin{figure}
\includegraphics[angle=0, width=0.99\columnwidth]{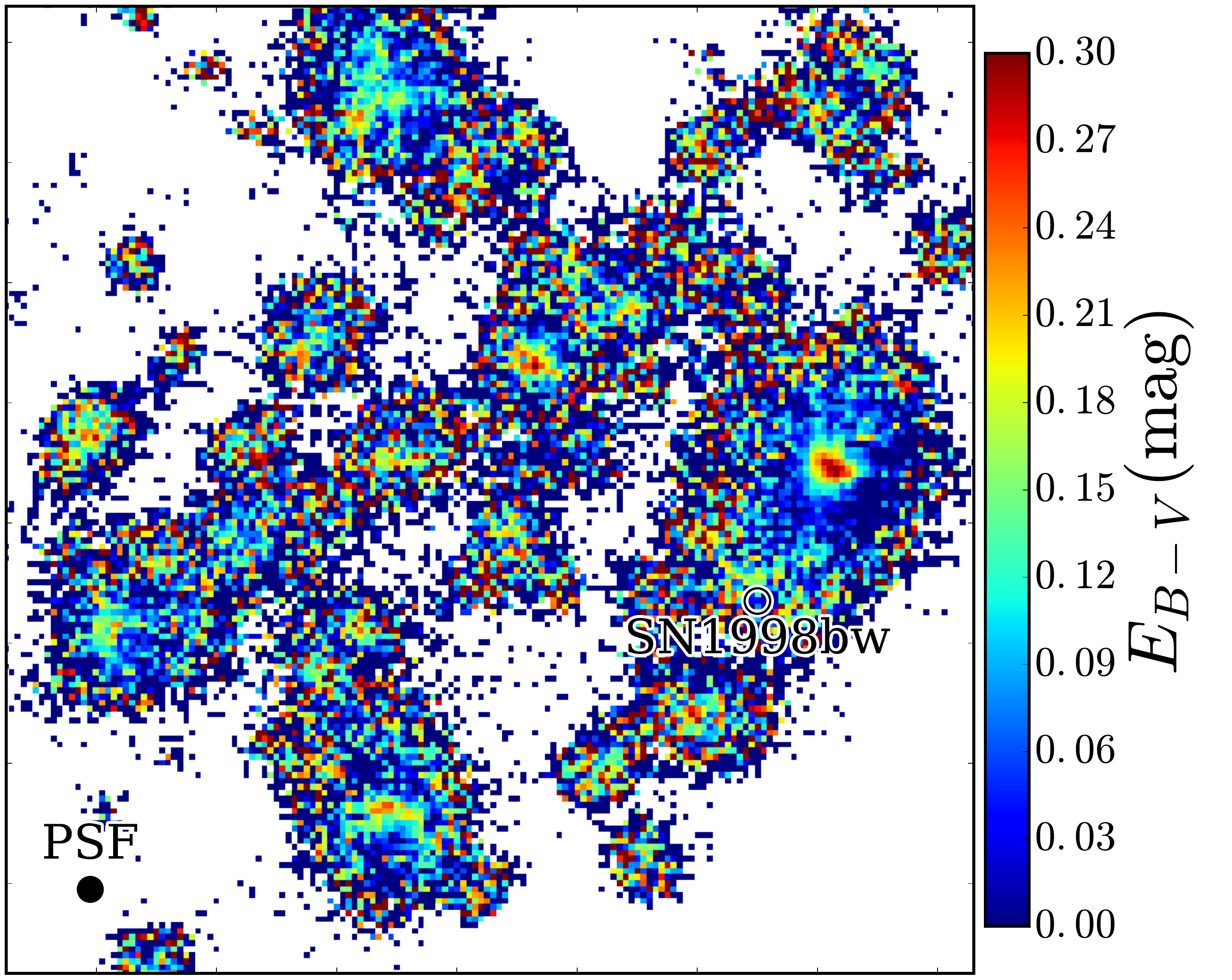}
\caption{Dust-reddening distribution in ESO184-G82 as measured through the Balmer decrement. We only show spaxels in which \hb~ is detected with a significance of at least $3\sigma$. The image spans 34\arcsec~by 38\arcsec, or 6.1~kpc by 6.8~kpc. The circle denoting the position of SN~1998bw has a {diameter} of 160~pc.}
\label{fig:ebv}
\end{figure}

Because SN~1998bw has an exquisite photometric and spectroscopic data coverage, it is widely used as a comparison object. Deriving its intrinsic properties such as luminosity or color, however, requires an accurate knowledge of the absorbing dust column in its host, which we infer here through Balmer lines. We convert line fluxes of \ha~and \hb~into a map of color excess $E_{B-V}$ as shown in Fig.~\ref{fig:ebv} using Eq. 5 of \citet{2015A&A...581A.125K}. This procedure assumes standard ratios of the Balmer lines from \citet{1989agna.book.....O} at a temperature or $10^4$~K and electron density $n_{\mathrm{e}}\sim100~\mathrm{cm}^{-3}$. These assumptions are broadly consistent with the values measured for ESO184-G82 (Table~\ref{tab:prop}). Our results depend only marginally on our choice of an average Milky Way extinction law \citep{1992ApJ...395..130P} with $R_V=3.08$, as the difference between reddening laws in the Local Group is small in the wavelength range probed by \ha~ and \hb. 

Figure~\ref{fig:ebv}~shows very little dust in general in ESO184-G82, and also only minor evidence for dust at the spaxel closest to the actual GRB or SN position ($E_{B-V} = 0.04\pm0.03~\mathrm{mag}$ or $A_V = 0.13\pm0.09~\mathrm{mag}$) and its immediate environment ($E_{B-V} = 0.03_{-0.03}^{+0.05}~\mathrm{mag}$ in the nine spaxels closest to the GRB position). The only locations where we observe significant dust reddening are the centers of \hii~ regions as exemplified by the WR region (Fig.~\ref{fig:ebv}). These areas show a centrally symmetric substructure in dust extinction that is decreasing from the inside out and peaking at $E_{B-V} \sim 0.25~\mathrm{mag}$ or $A_V \sim 0.8~\mathrm{mag}$. A galaxy-integrated spectrum yields $E_{B-V} = 0.05\pm0.02$~mag or $A_V=0.15\pm0.06$~mag, which is in remarkable agreement with the average optical depth $\tau_V$ derived from modeling the UV-to-radio spectral energy distribution \citep{2014A&A...562A..70M}.

These low dust reddening values are somewhat in tension with results from optical spectroscopy in previous works \citep{2006A&A...454..103H, 2008A&A...490...45C}. While we can reproduce our MUSE values with a re-reduction of the archival data used in \citet{2006A&A...454..103H}, as shown in Appendix~\ref{app:fors}, the origin of the mismatch to the spatially resolved data of \citet{2008A&A...490...45C} remains unclear\footnote{Potential reasons include uncertainties in the flux calibration or the stellar Balmer absorption correction; the disagreement in $A_V$ seems strongest where the \hb~fluxes are lowest.}.

The reliable spectrophotometric calibration of the MUSE data (Sect.~\ref{sec:red}), our accurate stellar continuum modeling (Sect.~\ref{sec:stargas}) and the consistency with the {spatially resolved dust properties from long-wavelength data \citep{2014A&A...562A..70M}} leads us to believe that our new data now resolve the apparent conflict between an unexpectedly large reddening at the SN position derived in previous {works} and the SN itself, which did not show strong evidence of dust obscuration \citep[e.g.,][]{1998Natur.395..672I, 2001ApJ...555..900P}. These data provide further confidence in using SN~1998bw as an only marginally reddened SN template for comparison to other events \citep[e.g.,][and references therein]{2004ApJ...609..952Z, 2014A&A...566A.102S, 2016arXiv160606791K}.

\subsection{Metallicity diagnostics}

\begin{figure*}
\centering
\begin{subfigure}{.42\textwidth}
  \includegraphics[width=0.999\linewidth]{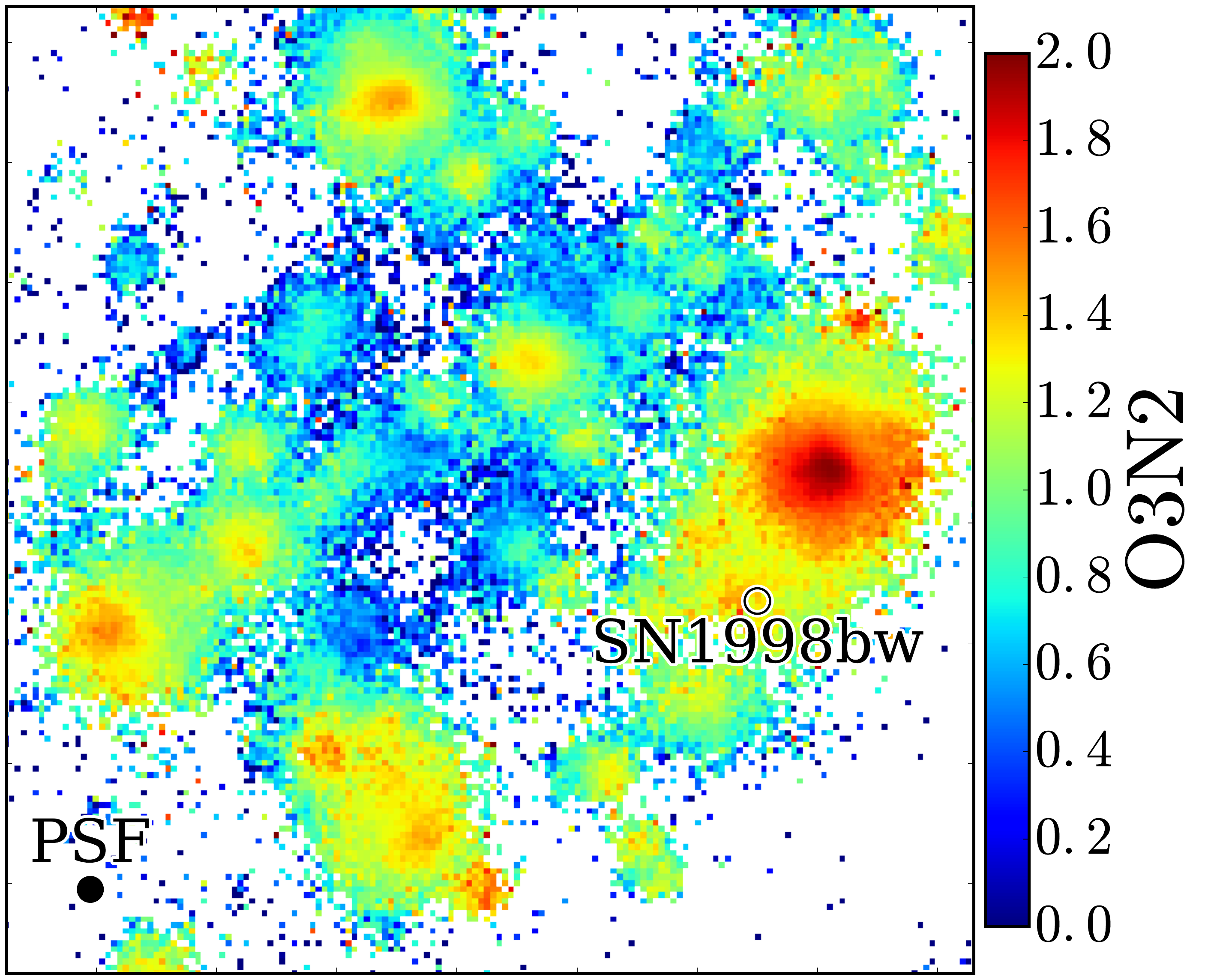}
\end{subfigure}
\begin{subfigure}{.42\textwidth}
  \includegraphics[width=0.999\linewidth]{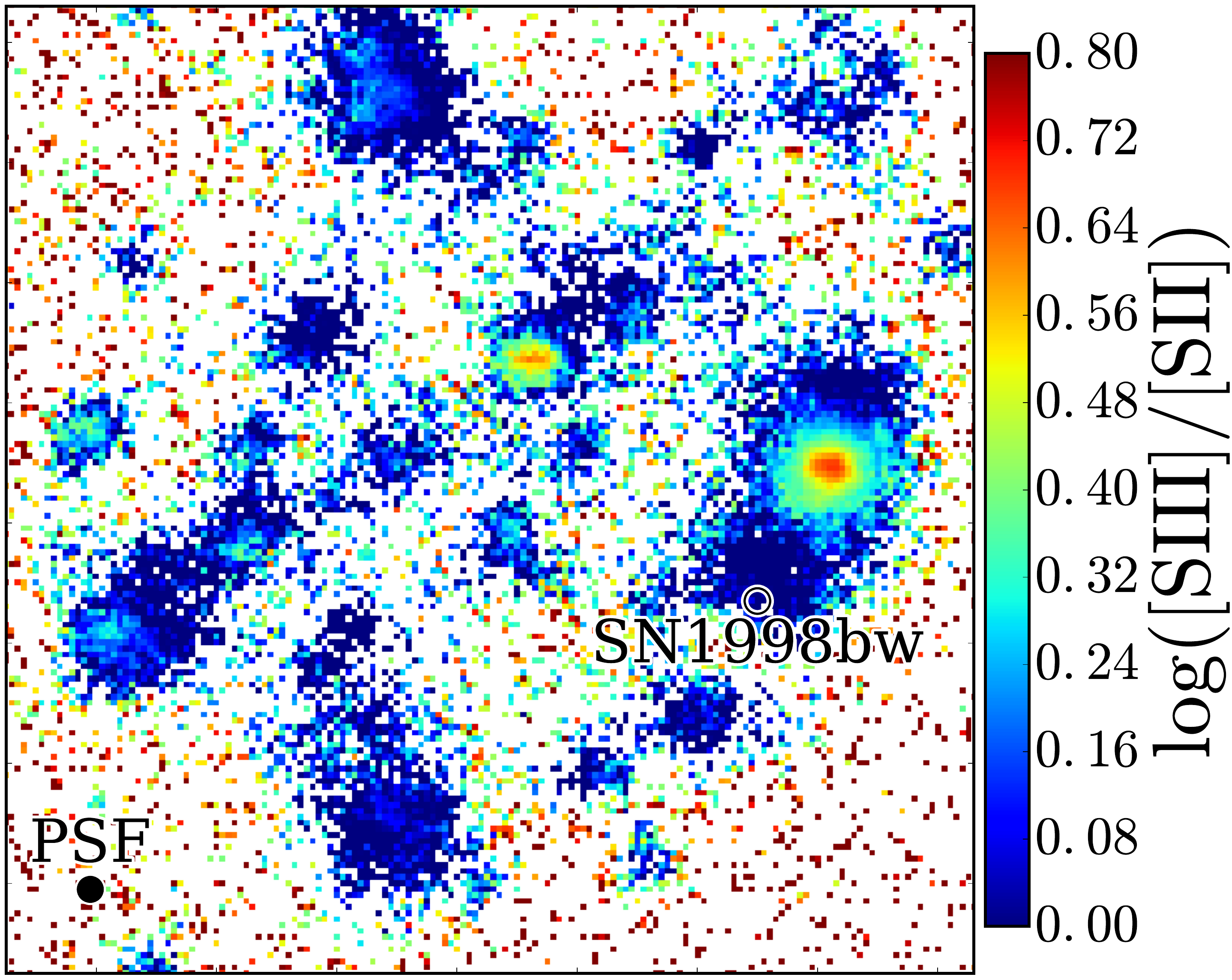}
\end{subfigure}
\caption{Face-to-face comparison between the ESO184-G82 maps of O3N2, often used for abundance determinations and \siii/\sii, a tracer of the ionization state of the hot gas. Only spaxels with {S/N > 3} are shown. Image dimensions are similar to Fig.~\ref{fig:ebv}.}
\label{fig:s3s2}
\end{figure*}

\subsubsection{Initial considerations}

Metal abundances of \hii\  regions are a central observable to study cosmochemical evolution, and a large set of literature is devoted to the various possibilities, their advantages, and perils to infer abundances from \hii-region spectra \citep[e.g.,][]{1979MNRAS.189...95P, 1991ApJ...380..140M, 2005ApJ...631..231P, 2008ApJ...681.1183K}. Very briefly, the most common methods to deduce chemical abundances, and from those, the abundance of oxygen (traditionally expressed in $\oh$) make use of either photoionization models \citep[e.g.,][]{1985ApJS...58..125E, 2000ApJ...542..224D, 2002ApJS..142...35K} or empirical correlations between certain strong-line ratios and oxygen abundances derived through electron temperatures $T_{\rm{e}}$ from collisionally excited lines \citep[CELs; e.g.,][]{2004MNRAS.348L..59P, 2013A&A...559A.114M}. Commonly used ratios are, for example, \nii/\oii, \oiii/\nii,  \nii/\ha, or $R_{23}$ = (\oii+\oiii)/\hb, which have been (re)calibrated numerous times against different samples of $T_{\rm{e}}$-based abundances or photoionization models, yielding a large set of different calibrators in the literature \citep[e.g.,][]{2002ApJS..142...35K, 2004ApJ...617..240K, 2005ApJ...631..231P, 2006A&A...459...85N, 2008A&A...488..463M}.

The most fundamental problem in using and interpreting the oxygen abundances derived in this way is that different methods are only very rarely consistent \citep[e.g.,][]{2008ApJ...681.1183K}, which gives rise to the abundance determination problem \citep{1967ApJ...150..825P}. Methods based on temperature-sensitive CELs typically show abundances that are lower by 0.2~--~0.4 dex with respect to photoionization-based methods or abundances derived with temperatures from recombination lines \citep[e.g.,][and references therein]{2012MNRAS.426.2630L}, in particular in the high-metallicity region. A possible solution to the abundance determination problem are small-scale temperature fluctuations \citep[e.g.,][]{2003ApJ...584..735P, 2004MNRAS.355..229E} or an electron population that is distributed somewhat differently than in thermal Maxwell-Boltzmann equilibrium \citep{2012ApJ...752..148N, 2012MNRAS.426.2630L}. But until these discrepancies are fully resolved, element abundances from emission lines remain the subject of large controversy.

A second, independent problem relates to the observational difficulties in robustly measuring emission-line fluxes for lines in different wavelength ranges for faint, high-redshift galaxies. As a consequence of various observational constraints, the available data are typically limited to a handful of strong lines. This is similarly true for our observations here, as the MUSE data do not cover the strong \oii($\lambda\lambda3726,3729$)~doublet nor \oiii($\lambda 4363$), which is one of the most commonly used, temperature-sensitive CEL. A very popular emission-line diagnostic in the literature has thus been the logarithm of the ratio of \oiii~/\hb~to \nii/\ha~ or short O3N2 \citep[e.g.,][]{2004MNRAS.348L..59P, 2013A&A...559A.114M} because of its independence on dust reddening and the relative observational ease with which it can be measured even at $z\sim 2$.

\subsubsection{Specific problems of empirical metallicity diagnostics}

From the very different ionization potentials (IPs) of N and O$^{+}$ (14.5~eV versus 35.1~eV), it is {directly} clear that O3N2 should also carry a strong dependence on the ionization parameter in addition to its inverse proportionality to oxygen abundance \citep[e.g.,][]{1979A&A....78..200A, 2015MNRAS.448.2030H}. In Fig.~\ref{fig:s3s2}, we plot the O3N2 map of ESO184-G82, which immediately translates into a map of oxygen abundance ($\oh$~would be lowest where O3N2 is highest) in common strong-line diagnostics \citep{2004MNRAS.348L..59P}.

When inspecting the left part of Fig.~\ref{fig:s3s2}, however, it becomes apparent that O3N2-based oxygen abundances (and similarly for those from N2) produce abundance maps that are hard to understand in a physical context: O3N2 varies significantly on $\lesssim\mathrm{kpc}$ scales leading to an unexpected\footnote{Despite the filamentary structure of nearby giant \hii~regions such as 30 Doradus, they are usually adequately described with abundances that are homogeneous throughout the region \citep[e.g.,][and references therein]{2011ApJ...738...34P}.} chemically inhomogeneous structure within individual \hii~regions with their central abundances up to 0.3 dex lower than their outer edges. However, we believe that the significant gradients in O3N2/N2 observed in most of the \hii~regions are unlikely to be due to a genuine variation in oxygen abundance, but are more likely to result from a changing ionization parameter. We explore this hypothesis further in the following sections.

\subsubsection{Ionization map}

Our MUSE data are of sufficient depth and quality to test how strongly O3N2 is affected by ionization empirically through the ratio of \siii~($\mathrm{IP}=23.3$~eV) to \sii~($\mathrm{IP}=10.3$~eV); this ratio is widely considered as one of the best tracers of the ionization parameter \citep{1991MNRAS.253..245D} as it shows, in contrast to \oiii/\oii, only very little dependence on abundance itself \citep{2002ApJS..142...35K, 2011MNRAS.415.3616D}. The resulting map\footnote{As MUSE does not cover the wavelength range of \siii($\lambda$9532), we use a theoretical value of $\siii(\lambda9532)=2.44\times\siii(\lambda9069)$ \citep{1982MNRAS.199.1025M} here.} is shown in Fig.~\ref{fig:s3s2} and clearly highlights that the center of the \hii~regions have the highest values of \siii/\sii, and thus ionization parameter.

This adds further support to our initial conjecture and ascertains that the radially symmetric structure of individual \hii\  regions in the O3N2 map is due to an increase of the ionization parameter toward the center and not a decrease in chemical abundance. Simple O3N2 (or N2-based) diagnostic line ratios are thus inadequate to produce accurate maps of oxygen abundance at the level of detail of our MUSE data (see also Appendix~\ref{sec:abundancevsion}). Comparing the specific measurements of oxygen abundance in Table~\ref{tab:prop}, it becomes clear that the O3N2-based diagnostic systematically underestimates $\oh$~for high ionization parameters as observed in the WR region, and overestimates $\oh$\ for regions of lower ionization such as the SN explosion site.

\subsubsection{Metallicity map}
\label{sec:mapoh}

\begin{figure}
\includegraphics[angle=0, width=0.99\columnwidth]{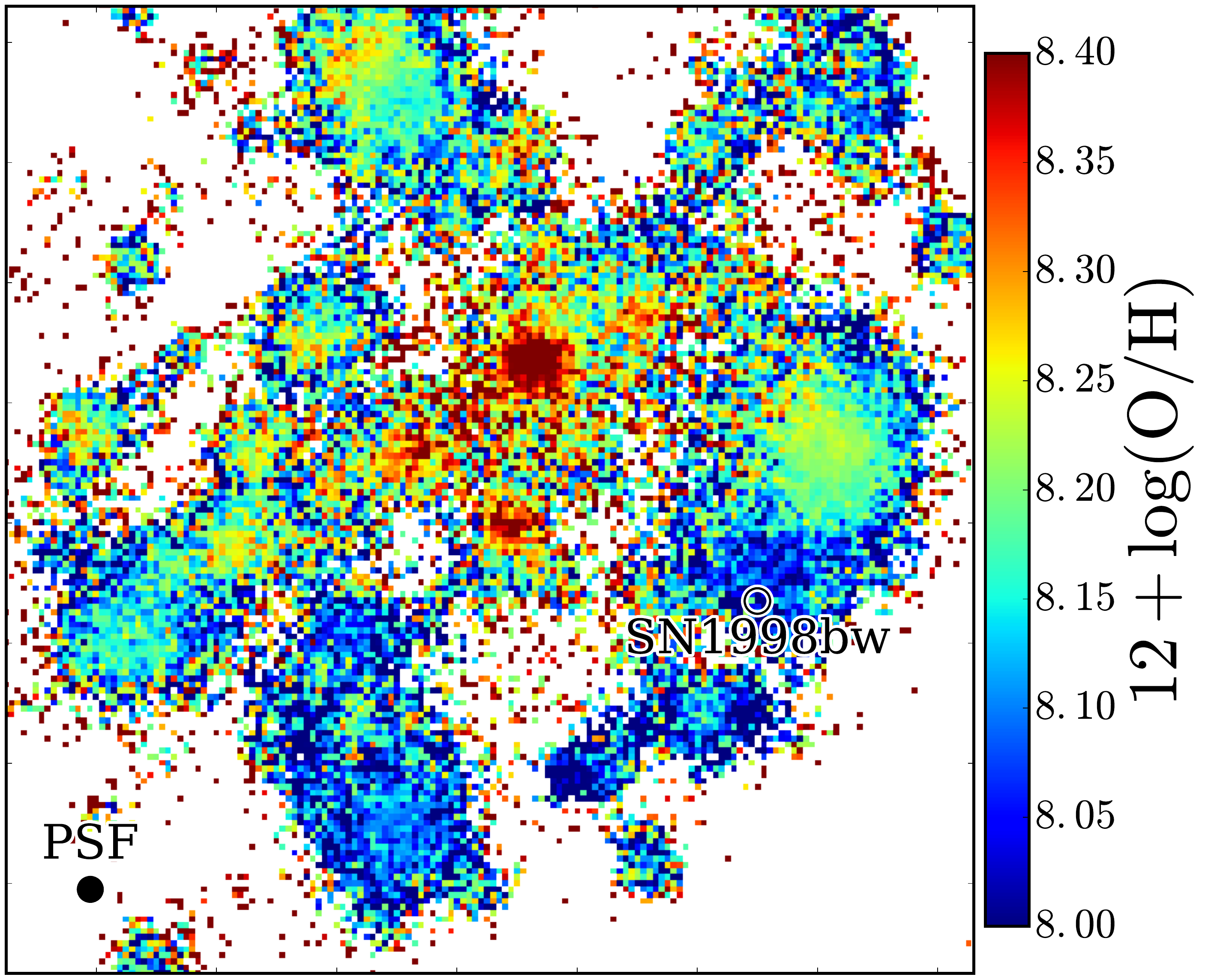}
\caption{Map of $\oh$~ as obtained through the \sii~ and \nii-based method of \citet{2016Ap&SS.361...61D}. Only spaxels with {S/N > 3} are shown. Image dimensions are 34\arcsec~by 38\arcsec or 6.1~kpc by 6.8~kpc, similar to Fig.~\ref{fig:ebv}. Methods to estimate $\oh$ based on electron temperatures return values that are 0.1~--~0.2 dex higher.}
\label{fig:s2}
\end{figure}

After rejecting common empirical methods using O3N2 or N2 as accurate metallicity tracers due to their ionization dependence, we turn to diagnostics based on photoionization models. Unfortunately, most of the previous strong-line methods rely in one way or another on the strong \oii($\lambda\lambda$3726,3729)~ doublet \citep{2002ApJS..142...35K}, which is not directly available to us here. Also \siii/\sii~ is a good ionization tracer, but \siii~is relatively faint and not detected in most of our spaxels (see Fig.~\ref{fig:s3s2}). 

A recently published method based on photoionization modeling \citep{2016Ap&SS.361...61D} seems to perfectly fit to our data; it relies solely on \ha, \nii, and \sii, which are all strong and well within the wavelength range of MUSE. The method, shortened as D16 in the following, is introduced as "effectively independent of both ionization parameter and ISM pressure" \citep{2016Ap&SS.361...61D};  Fig.~\ref{fig:s2} shows the respective map of $\oh$~ in regions where all necessary lines are detected at a S/N of at least 3.

Clearly, the strong abundance gradient over individual \hii\  regions, as would have been deduced from O3N2, is not observed in this diagnostic. Instead, the oxygen abundance map shows a relatively smooth behavior with a decreasing overall metallicity from the center of the galaxy toward the outside (see also Sect. \ref{sec:metgrad}). The spaxel abundance at the SN position is $\oh=8.00$ or 0.20~$Z_{\odot}$. The immediate environment is consistent with this value and homogeneous: spaxels within a radius of 70~pc to the SN position yield $\oh = 8.04\pm 0.06$. The WR region displays a somewhat higher oxygen abundance with $\oh=8.22\pm 0.03$ or $0.34\pm0.03~Z_{\odot}$ around the peak of the \ha~emission.

It is of course reasonable to ask now whether the new D16 diagnostic provides more reliable constraints on oxygen abundance than previous methods given the significant differences that exists between all of them \citep{2016arXiv161108595B}. In addition, for low-mass galaxies as is the case here, this diagnostic seems to return lower oxygen abundances than previous methods \citep{2016ApJ...823L..24K}. To elaborate further on the \sii-based diagnostic, we reproduce the combined Eq. 1 and 2 from \citet{2016Ap&SS.361...61D}, i.e.,

\begin{equation}
\label{eq:s2}
12+\log(\mathrm{O/H}) = 8.77 + \log([\ion{N}{ii}]/[\ion{S}{ii}]) + 0.264\log([\ion{N}{ii}]/\mathrm{H}\alpha)
,\end{equation}

where \nii~ is the flux in the \nii($\lambda6484$) line and \sii~ the flux in the \sii($\lambda\lambda6717,6731$) doublet. The primary observable is thus the nitrogen to sulfur ratio, which is a tracer of the nitrogen to oxygen ratio\footnote{Sulfur and oxygen are both $\alpha$-process elements produced in massive stars and observed to track each other well in different environments \citep[see, e.g., Fig. 6 in][]{2006A&A...448..955I}.}. Because nitrogen is also produced in intermediate-mass stars, N/O starts to depend on $\oh$~above $\oh\sim 7.8$ \citep[e.g.,][]{1999ApJ...511..639I, 2013A&A...549A..25P, 2016A&A...595A..62P}, and as expected, a N/O map via \citet{2010ApJ...715L.128A} is very similar to the map of oxygen abundance in the respective strong-line diagnostic.

The map of oxygen abundance then fundamentally relies on the N/O-to-O/H calibration, and it is in principle not impossible that the applied calibration is slightly offset, in particular in the low-metallicity region. However, the calibration sample for N/O-to-O/H in the metallicity range of interest is based on low-metallicity blue compact dwarf galaxies \citep{1999ApJ...511..639I}, which are not dissimilar in physical properties to our {galaxy}. In addition, there could also be internal variations in N/O at a given $\oh$ or differences between the host of SN~1998bw to the calibration sample. For example, infall of primordial gas \citep[e.g.,][]{2015A&A...582A..78M} would decrease $\oh$, but leave N/O unaffected \citep{2016ApJ...823L..24K}. Another point of concern would be an anomalously high N/O ratio for the SN region as claimed by \citet{2006A&A...454..103H}. Both of these effects would lead us to overpredict the actual oxygen abundance in the SN region via the D16 diagnostic. {The oxygen abundances derived through Eq.~\ref{eq:s2}, however, are already lower than the values from temperature-sensitive collisionally excited lines (Sect.~\ref{teoh}), which are sensitive to the actual oxygen abundance. We hence find no evidence for a systematic overestimation of the metallicity in the D16 scale at either the explosion site or the WR region.}

\subsubsection{Metallicities based on electron temperatures}
\label{teoh}

\begin{figure}
\begin{subfigure}{.24\textwidth}
\includegraphics[width=0.999\linewidth]{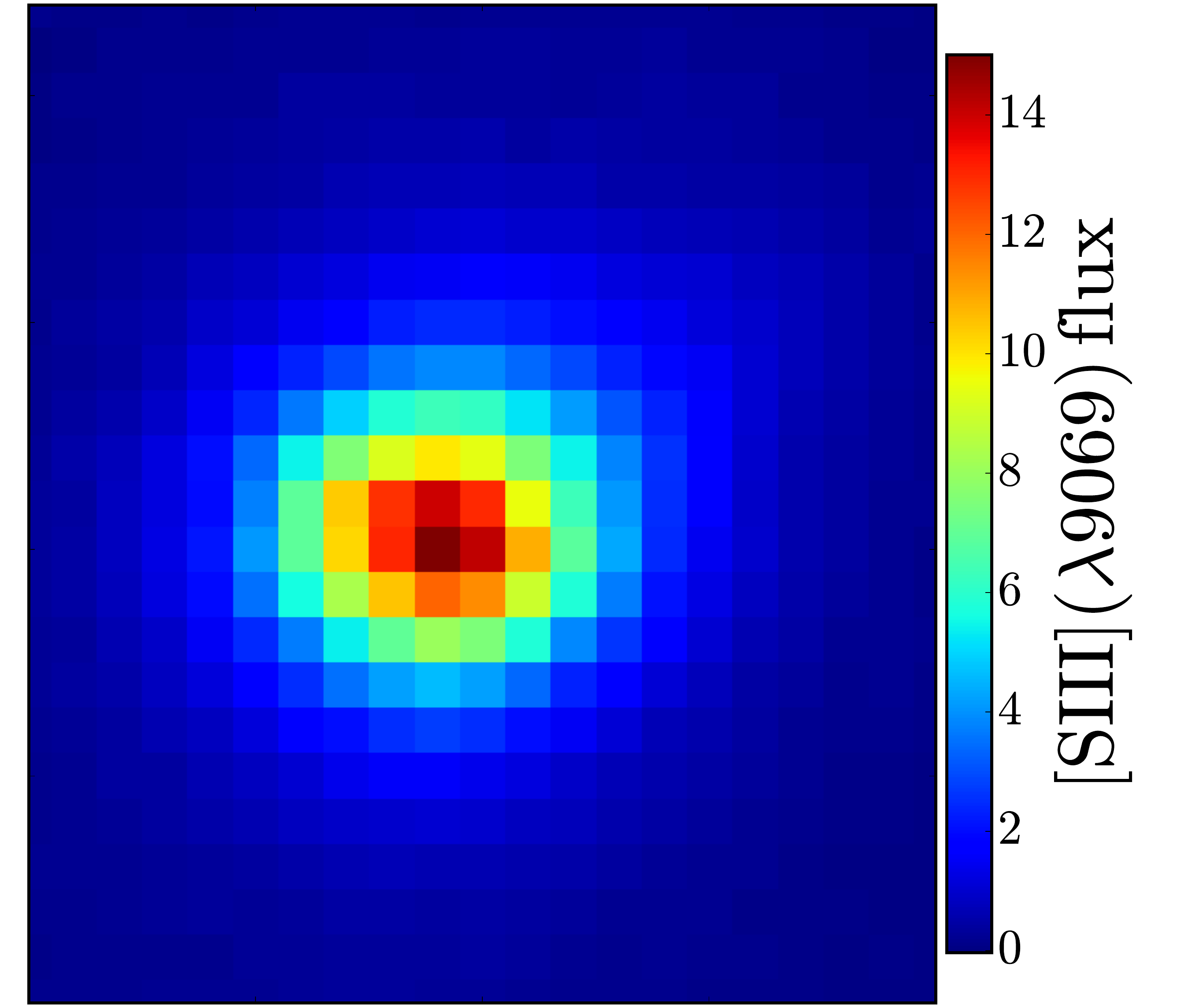}
\end{subfigure}
\begin{subfigure}{.24\textwidth}
\includegraphics[width=0.999\linewidth]{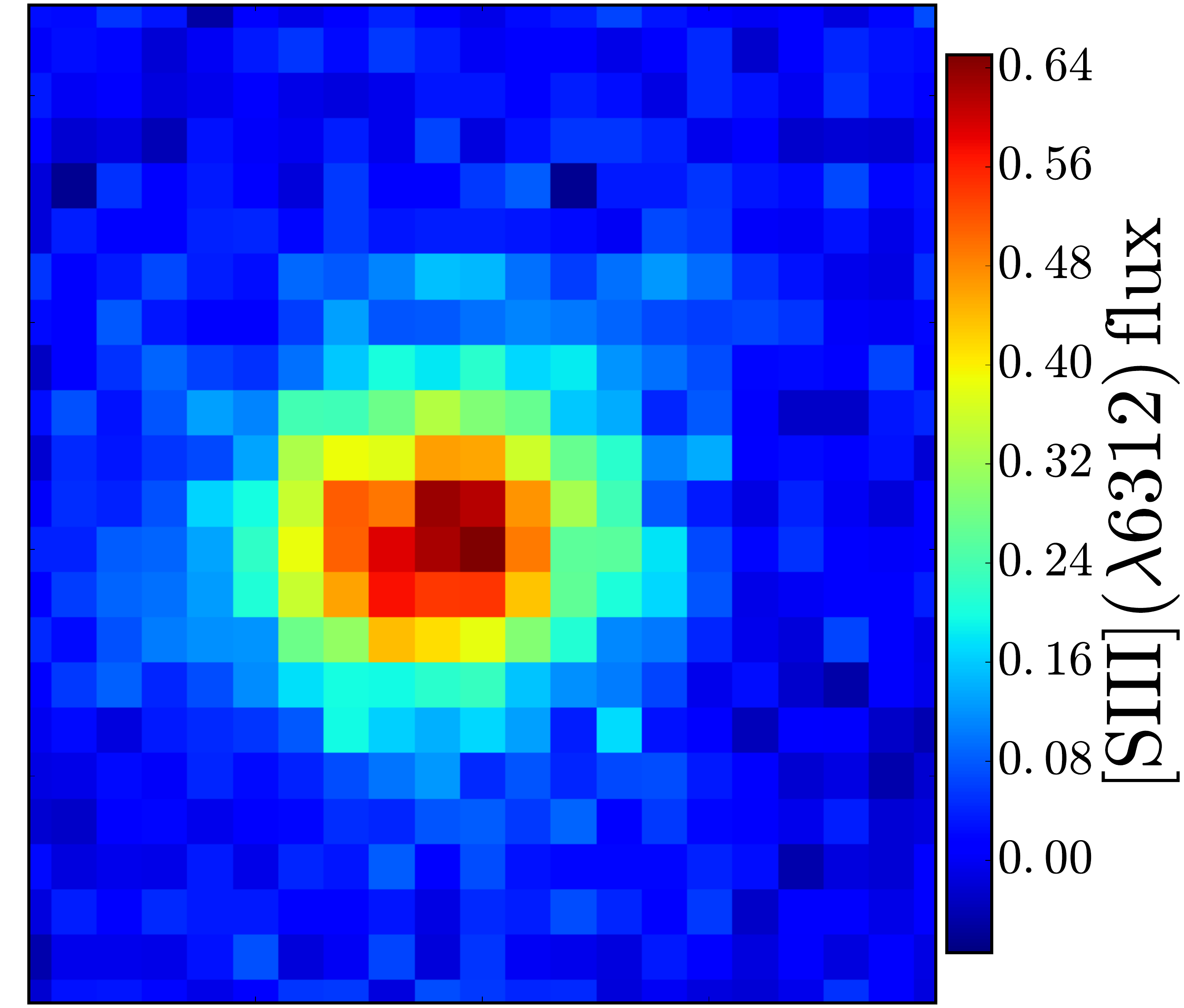}
\end{subfigure}
\begin{subfigure}{.24\textwidth}
\includegraphics[width=0.999\linewidth]{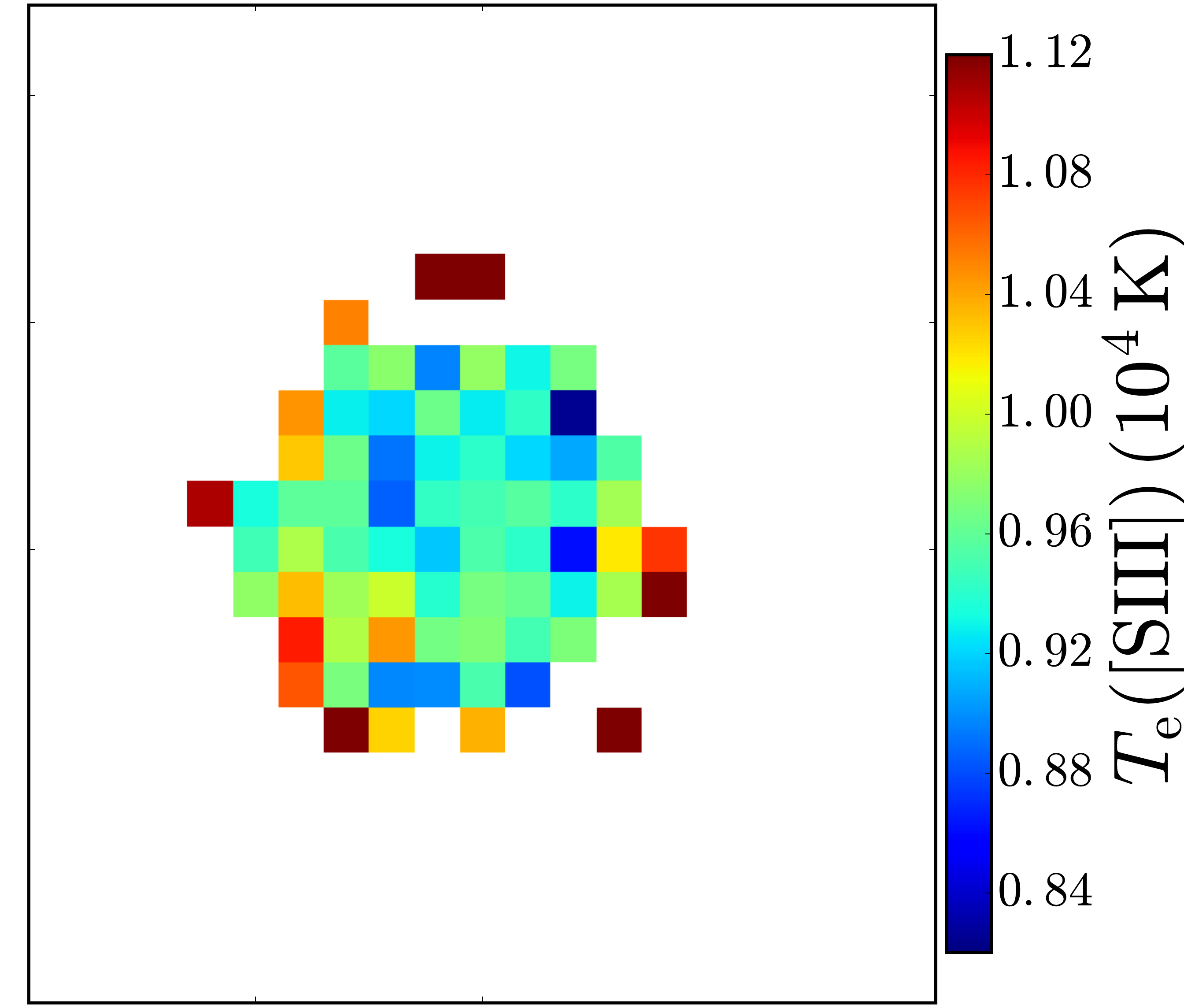}
\end{subfigure}
\hspace{0.05cm}
\begin{subfigure}{.24\textwidth}
\includegraphics[width=0.999\linewidth]{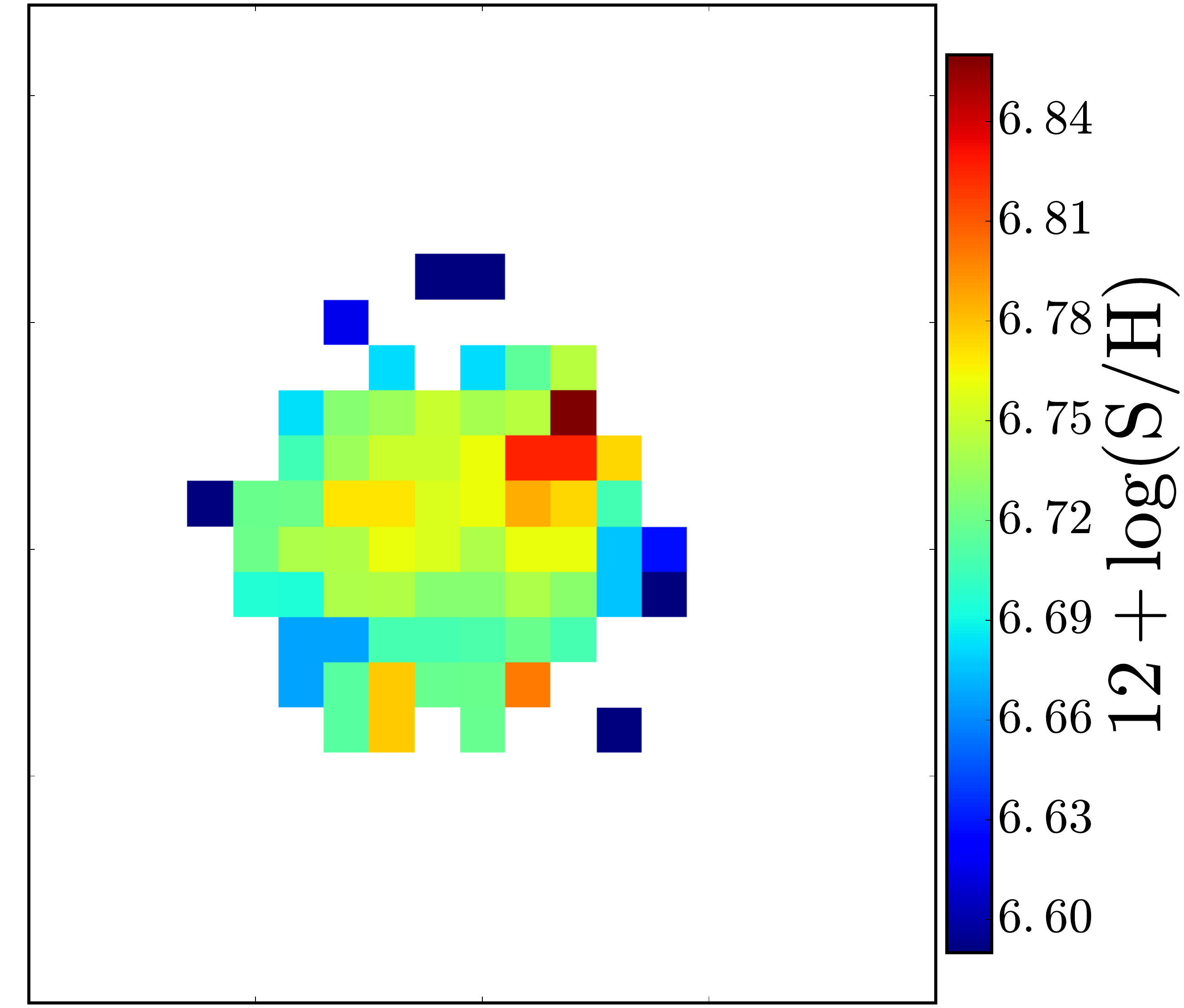}
\end{subfigure}

\caption{Electron temperatures ($T_{\mathrm{e}}$) from \siii\ and corresponding sulfur abundances in the WR region. \textit{Top left}: flux map of nebular \siii($\lambda9069$). \textit{Top right}: flux map of auroral \siii($\lambda6312$). Both flux maps are in units of $10^{-17}$\,\erg. \textit{Bottom left}: electron temperatures from \siii. \textit{Bottom right}: $12+\log(\mathrm{S/H})$ as derived from $T_{\mathrm{e}}$. The solar abundance [O/S] is 1.57 \citep{2009ARA&A..47..481A}, so the $12+\log(\mathrm{S/H})$ scale corresponds to $\oh = 8.2$ to 8.4. All panels are approximately 6\arcsec~by 6\arcsec, or 1~kpc by 1~kpc.}
\label{fig:temp}
\end{figure}

Given the strong constraints that these measurements of oxygen abundance could imply for the SN~1998bw progenitor metallicity, we further seek to corroborate our earlier abundances through those from temperature-sensitive lines. Unfortunately, the brightest and most commonly used auroral line \oiii($\lambda$4363) is not covered by the MUSE wavelength response and the other temperature-sensitive lines are too faint to be detected in most of the individual spaxels. However, we clearly detect \siii($\lambda$6312) both in the WR region (Fig.~\ref{fig:temp}) and in integrated spectra around the SN position.

Following \citet{2013ApJS..207...21N}, we derive an electron temperature $T_\mathrm{e}$ using \siii~ in the central part of the WR region of around $T_{\mathrm{e}}({\siii})=(0.94\pm0.02)\cdot10^{4}$~K. Figure~\ref{fig:temp} contains maps of the crucial emission lines, i.e., nebular \siii($\lambda9069$) and auroral \siii($\lambda$6312); the resulting temperatures; and sulfur abundances. These values correspond to somewhat larger temperatures of \oiii~of around $T_{\mathrm{e}}(\oiii) \sim 1.07\cdot10^{4}$~K \citep{2006A&A...448..955I, 2012A&A...547A..29B}. The flux in the doublets of \oii($\lambda\lambda$7320,7330), \oiii($\lambda\lambda$4959,5007) and \hb~ then yields a central abundance of the WR region around $\oh \sim 8.3$.

Similar values are obtained when using solar abundances to convert the measured sulfur to an oxygen abundance. These are slightly (0.1~--~0.2~dex) higher than those implied by the strong-line diagnostic from Sect. \ref{sec:mapoh}, but critically depend on the sulfur-to-oxygen temperature conversion or assumed abundance. They are thus subject to some systematic uncertainties. However, it is clear that the strong structure in O3N2 or N2 over the \hii\  region are not observed in electron temperatures.

Emission lines at the explosion site are substantially fainter and \siii($\lambda$6312) is only detected in a stacked spectrum extracted from three by three spaxels around the SN position, which yields $T_{\mathrm{e}}(\siii)=(1.24\pm0.18)\cdot10^{4}$~K. This is slightly higher than in the WR region, but with large uncertainties. The corresponding oxygen abundance derived in a similar manner as above is $\oh=7.92\pm0.27$.

We further use here our reduction of the public archival VLT long-slit spectroscopy (Appendix~\ref{app:fors}) as it covers both the WR region and SN position and extends below 4000~\AA. From the well-detected \oiii($\lambda$4363) line, we measure temperatures of $T_{\mathrm{e}}(\oiii)=(1.05\pm0.05)\cdot 10^{4}$~K and $T_{\mathrm{e}}(\oiii)=(1.29\pm0.18)\cdot10^{4}$~K for the WR and SN region, which are broadly consistent with the estimates from MUSE through \siii($\lambda$6312). They imply oxygen abundances of $\oh=8.36\pm0.07$ and $8.09\pm0.15$ for WR and SN region, respectively.

Our relatively low oxygen abundances at the location of explosion site, WR region, as well as the galaxy-integrated value ($\oh\sim8.3$; Sect.~\ref{sec:int}) correspond to 0.2 to 0.5 times the solar value, and thus explain many of the long-wavelength properties of ESO184-G82 as typical for metal-poor dwarf galaxies without invoking a deficiency in molecular gas \citep{2016arXiv160901742M}.

\subsection{Metallicity gradient}
\label{sec:metgrad}

The metallicity of galaxies is often observed to decrease with the distance from their centers \citep[e.g.,][]{1994ApJ...420...87Z, 2014A&A...563A..49S}, which is also seen in the hosts of SNe \citep{2016A&A...591A..48G} and here (Fig.~\ref{fig:s2}). These gradients are important to understand for spatially unresolved studies at high redshift, where positional offsets can be measured, but abundances are only derived in a galaxy-integrated manner. 

\begin{figure}
\includegraphics[angle=0, width=0.99\columnwidth]{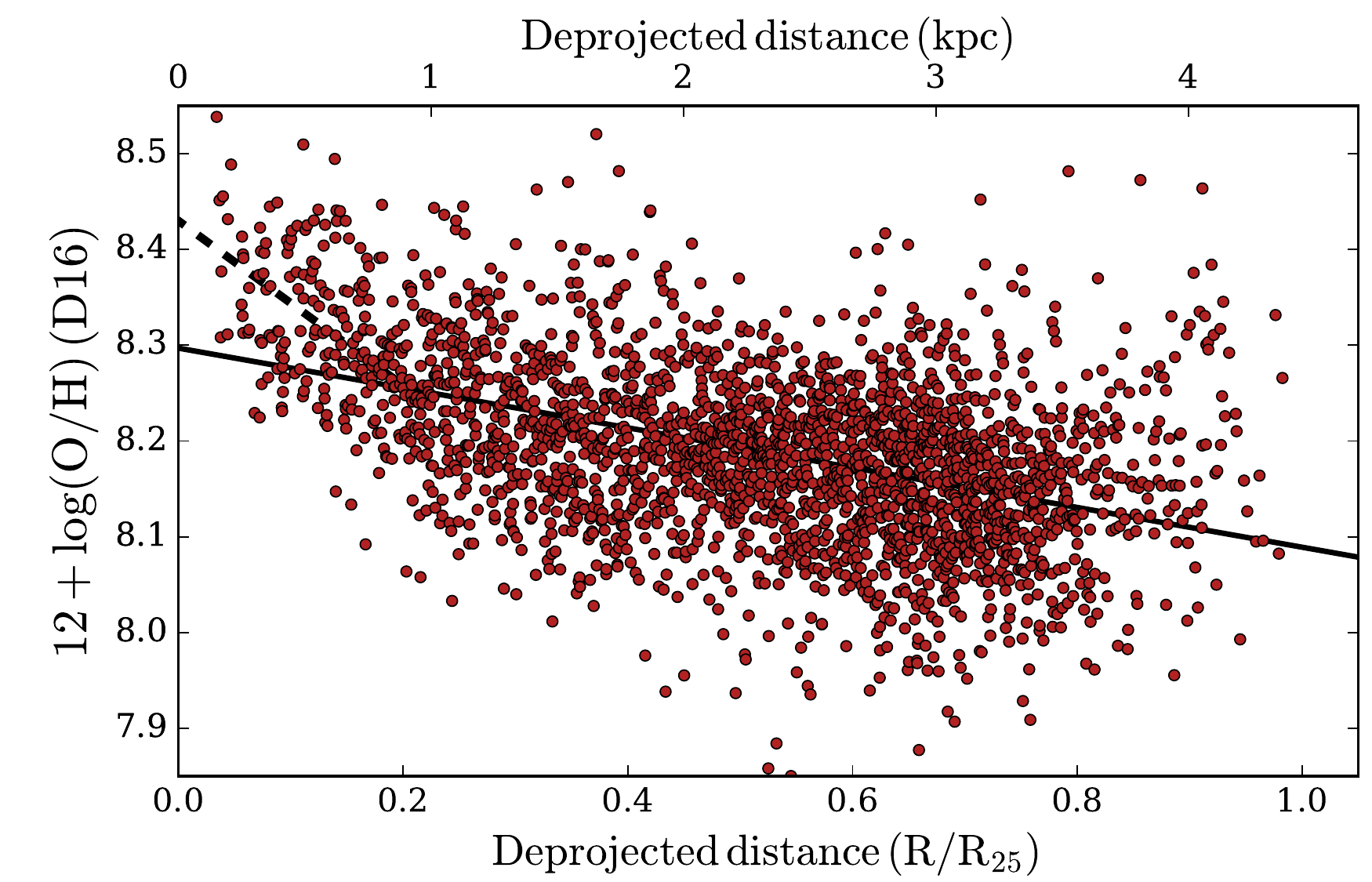}
\caption{Metallicity gradient for ESO184-G82 in  the D16 scale. Each data point is a spaxel-based measurement of oxygen abundance, where individual spaxels were binned with their direct neighbors to enhance the S/N if necessary.}
\label{fig:metgrad}
\end{figure}

Using a linear regression on the data in Fig.~\ref{fig:metgrad}, the metallicity gradient is best fit with a slope of $-0.25\pm0.02$~dex/$R_{25}$ in relative, or $-0.06$~dex~kpc$^{-1}$ in physical scales, which is well in the range that was previously reported for galaxies of comparable stellar mass \citep{2015MNRAS.448.2030H}. Here, $R_{25}$ is the radius at the $B=25~\mathrm{mag}~\mathrm{arcsec}^2$ isophote. The O3N2-based diagnostics return slightly steeper, but generally compatible values within
errors. The linear fit to the oxygen abundance data is a reasonable description of the data, except for the very center (deprojected distance < 1~kpc), where the metallicity is seen to decrease more steeply. Limiting the fit range to this region, the slope in the central kpc is $-0.18$~dex~kpc$^{-1}$ (see dashed line in Fig.~\ref{fig:metgrad}), but it depends strongly on the metallicity diagnostic\footnote{A O3N2-based method returns a decreasing abundance toward the center, similar to what is seen also in a third of SN hosts in this diagnostic \citep{2016A&A...591A..48G}.}.

A comparison between this metallicity gradient to the typical values of (projected) GRB distances from the galaxy centers\footnote{{GRBs are commonly found in dwarf galaxies with an irregular morphology \citep[e.g.,][]{2006Natur.441..463F, 2017MNRAS.tmp..220L}. Galaxy centers and thus positional offsets are hence sometimes difficult to determine robustly.}} of 1.3~kpc \citep{2016ApJ...817..144B} does not provide strong reason to suggest that the average measurement of GRB host metallicities from spatially unresolved data is significantly skewed when compared to the GRB site metallicity. There is, however, a non-negligible fraction of GRBs at substantial distances to their hosts (10\% of GRBs are located at offsets $>3$~kpc) where metallicity gradients might lead to overestimates of the GRB site abundance from unresolved spectra for single objects.

\subsection{ESO184-G82 if seen at high redshift}
\label{sec:int}

{GRB~980425 is the closest GRB since the discovery of GRB afterglows} and provides an exceptional opportunity to measure the physical parameters of its host with high precision and high spatial resolution. Typically, information on the environments of GRBs or similar kinds of transients at higher redshift has only been obtained through galaxy-integrated measurements \citep{2015A&A...581A.125K, 2016A&A...590A.129J, 2017A&A...599A.120V}, where it is not necessarily obvious how well the measured parameters actually correspond to properties at the GRB location. {We thus compare the physical parameters derived from the explosion site spectrum and the WR region, extracted from the nine spaxels closest to the SN position and the WR region center, to the parameters obtained from a galaxy-integrated spectrum of ESO184-G82. The latter is a simple sum of all spaxels within a radius of 25\arcsec\ around the center of the galaxy, which should mimic how the galaxy would appear if it were unresolved and at high redshift.}

First, we measure the total star formation rate of ESO184-G82 by adding up the \ha~flux of all individual spaxels after the respective spaxel-based reddening correction, which yields $SFR=0.22\pm0.02$~\Msunyr. If we instead use the \ha~flux from the galaxy-integrated spectrum and then apply a single reddening correction, we derive $SFR=0.17\pm0.02$~\Msunyr. Both values broadly agree with far-infrared, \oi, and \cii-based SFRs \citep{2014A&A...562A..70M, 2016arXiv160901742M}, and the narrowband \ha~image from \citet{2005NewA...11..103S} once their Salpeter IMF is taken into account \citep[see also][]{2009ApJ...691..182S}. The difference from other \ha-based SFR measurements \citep{2006A&A...454..103H, 2008A&A...490...45C} is entirely due to either an overestimated dust correction or a different IMF.

We then obtain further physical parameters from the extracted spectra of the explosion site, the central part of the Wolf-Rayet region, and the integrated-galaxy spectrum as summarized in Table~\ref{tab:prop}. Generally, there is decent agreement between many of the physical properties of the galaxy and the SN site spectrum. Not unexpectedly, resolved measurements of EWs are significantly higher, whereas both dust reddening (by 0.03~mag) and oxygen abundance (by 0.1~--~0.2~dex) are marginally lower at the explosion site than for a galaxy-integrated spectrum. However, ESO184-G82 does not provide strong evidence that GRB position spectra are markedly different from galaxy-integrated values except for the obvious mismatch in equivalent width measurements. Whether this observation remains valid for a larger sample of GRB hosts remains to be seen, of course.

Comparing Table~\ref{tab:prop} with Fig.~\ref{fig:s2}~and Fig.~\ref{fig:metgrad} also illustrates that the oxygen abundance derived from a galaxy-integrated spectrum is not the central abundance of the galaxy \citep[see also, e.g.,][]{2016A&A...591A..48G}. As the metallicity determination in an unresolved case is a SFR-weighted measurement, it is dominated by \hii\  regions that are primarily located somewhat offset from the center, as illustrated in Fig.~\ref{fig:EW}, and thus corresponds to a measurement at a distance of around 2~kpc to 3~kpc from the center of ESO184-G82. 

{In general, GRB positions tend to be highly concentrated on the UV-brightest regions of their hosts \citep[e.g.,][]{2010MNRAS.405...57S, 2016ApJ...817..144B, 2017MNRAS.tmp..220L}. These are the same regions that dominate the nebular-line emission in an unresolved, SFR-weighted spectrum from which physical parameters such as metallicity or dust content are typically inferred at higher redshift.}

\begin{table*}[!ht]
\caption{Physical properties inferred from MUSE IFU spectroscopy for the GRB~980425 host galaxy ESO184-G82 \label{tab:prop}}
\centering
\begin{tabular}{cccc}
\hline
\hline\noalign{\smallskip}
 & {ESO184-G82} & {SN region} & {WR region} \\
\hline\noalign{\smallskip}

EW(\ha) (\AA)       & $56.0\pm1.4$ & $92\pm15$  & $>900$ \\
EW(\oiii($\lambda$5007))  (\AA)    & $34.5\pm0.9$ & $54\pm5$  & $>850$ \\
$E_{B-V}$ (mag)    & $0.06\pm0.02$ & $0.03_{-0.03}^{+0.05}$ & $0.22\pm0.02$ \\
$n_{\mathrm{e}}$ (cm$^{-3}$)$^{\mathrm{(a)}}$ & $80\pm10$ & $90\pm10$ & $150\pm10$  \\
$T_{\mathrm{e}}$(\siii) ($10^4$~K)$^{\mathrm{(b)}}$ & $\cdots$ & $1.24\pm 0.18$ & $0.94\pm 0.02$\\
$T_{\mathrm{e}}$(\oiii) ($10^4$~K)$^{\mathrm{(b,c)}}$ & $\cdots$ & $1.29\pm 0.18$ & $1.05\pm 0.02$\\
$\oh$~ (D16)$^{\mathrm{(b)}}$ & $8.13 \pm 0.01$ & $8.04 \pm 0.06$ & $8.22 \pm 0.03$ \\
$\oh$~ (PP04, O3N2)$^{\mathrm{(b)}}$ & $8.31 \pm 0.01$ & $8.31 \pm0.01$ & $8.11 \pm 0.01$ \\
$\oh$~ (PP04, N2)$^{\mathrm{(b)}}$ & $8.33 \pm 0.01$ & $8.32 \pm0.01$ & $8.18 \pm 0.01$ \\
$\oh$~ ($T_{\mathrm{e}}$)$^{\mathrm{(c)}}$ & $\cdots$ &  $8.09\pm0.15$ & $8.36\pm0.07$ \\
SFR (\Msunyr) & $0.22\pm0.02$ & $\cdots$ & $\cdots$ \\

\hline\noalign{\smallskip}
\end{tabular}

\tablefoot{
\tablefoottext{a}{Derived from the \sii($\lambda\lambda6716,6731$) doublet ratio.}
\tablefoottext{b}{The quoted error is statistical only. There is an additional systematic error in each of these measurements for various reasons. For the respective parameters, the errors are approximately $T_{\mathrm{e}}=0.1\cdot10^4$~K, $\oh$~ (PP04, O3N2)=0.14~dex, $\oh$~ (PP04, N2)=0.18~dex. The systematic error on $\oh$~ (D16) is presently not well quantified, but we expect that it is in the range of $\sim$0.1~dex.}
\tablefoottext{c}{Partially derived from the archival VLT long-slit spectra (Appendix~\ref{app:fors}).}

} 
\end{table*}

\section{Conclusions and summary}

The first solid observational evidence that linked GRBs with core-collapse supernovae was provided by SN~1998bw, and since then this object has become prototypical for SNe following GRBs in terms of luminosity, temporal evolution and spectral properties. In particular the lack of hydrogen and helium and the presence of broad metal absorption lines, characteristic of a photosphere expanding at high velocities, have been observed in most GRB-SNe \citep{2012grbu.book..169H, 2016LPICo1962.4116C}. 

In this article, we use spatially resolved spectroscopy obtained with the novel integral field unit MUSE at the VLT to study the properties of the hot gas phase and the stellar population in the SN~1998bw host galaxy with a particular {emphasis} on the physical parameters at the SN explosion site. Our data cover the largest part of the host galaxy with individual spaxels of size $35 \times 35$~pc. The effective spatial resolution is limited by the atmospheric conditions and characterized by the point spread function with a full width half maximum of {$160$~pc}. The main results derived from our analysis can be summarized as follows:

\textit{(i)} GRB~980425/SN~1998bw exploded in a young (5~--~8 Myr) and dust-poor ($E_{B-V} = 0.03_{-0.03}^{+0.05}~\mathrm{mag}$) environment. The age of the stellar population corresponds to lifetimes of stars with $M_{\rm{ZAMS}}$  between approximately 25~\Msun and 40~\Msun. Our measurement of dust reddening for the explosion site is significantly lower than previous estimates and resolves the mismatch between an apparently dust-rich environment and a non-extinguished SN light curve.

\textit{(ii)} The Wolf-Rayet region is located at a projected distance of 860~pc to the GRB explosion site and is extremely young (< 3 Myr). This young age would imply implausibly high peculiar velocities for runaway stars and makes scenarios where the GRB progenitor is ejected from the WR region very contrived. The progenitor mass obtained from modeling the nebular spectra of SN~1998bw \citep{2006ApJ...640..854M} is consistent with that derived from the explosion environment and is inconsistent with an origin in the WR region. This strongly suggests that the GRB formed in situ.

\textit{(iii)} The total, dust-corrected SFR of ESO184-G82 is $SFR=0.22\pm0.02$~\Msunyr inferred from integrating the dust-corrected \ha~flux in the respective spaxels. The galaxy has a metallicity gradient of $-0.06~\mathrm{dex~kpc^{-1}}$, suggesting that the typical offsets of at most few kpc between GRBs and their host center have a relatively small impact on the abundance determination for higher redshift GRB sites.

\textit{(iv)} Most physical parameters that are derived from emission lines are dominated by the most star-forming regions. An integrated spectrum of the full galaxy and can thus lead to an adequate representation of the actual dust extinction or oxygen abundance at the GRB location. A noteworthy exception are equivalent widths, where an integrated spectrum significantly underestimates the values from the explosion site.

Despite considerable systematic uncertainty stemming from the validity and accuracy of specific strong-line metallicity diagnostics, we reach here the following robust conclusions with respect to abundances:

\textit{(v)} Empirical strong-line methods using \oiii/\hb~and/or \nii/\ha~ fail to produce accurate maps of metallicity at the level of detail probed by our MUSE observation. They significantly under- or overestimate $\oh$~in regions of high or low ionization parameter, respectively, and therefore return an unphysical radial gradient in individual \hii\  regions with their centers appearing artificially deprived in metals.

\textit{(vi)} A recent method based on photoionization models and \sii~from \citet{2016Ap&SS.361...61D} does not show an obvious dependence of inferred abundance on ionization and returns values that are broadly similar, but somewhat lower (by 0.1~dex to 0.2~dex) than those from electron temperatures via the observed ratios of auroral-to-nebular \siii~or \oiii. 

\textit{(vii)} Taking the {above} considerations into account, we consider {$\oh\sim8.2$} at the SN position, $\oh\sim8.3$ for a galaxy-integrated spectrum, and $\oh\sim8.4$ for the nearby WR region as our best estimates of the respective oxygen abundances\footnote{{These values are based on the \citet{2016Ap&SS.361...61D} diagnostic, but scaled upward by 0.15~dex as indicated by the corresponding $T_{\mathrm{e}}$ abundances.}}.

The immediate environment of GRB~980425 thus indicates a progenitor with $Z\sim0.3~Z_\odot$ and $M_{\rm{ZAMS}}$ between $\sim25~M_{\odot}$ and $\sim40~M_{\odot}$. Despite the preeminent efficiency, sensitivity, and optical quality of MUSE, similar studies for a larger sample of GRB hosts remain observationally challenging as a result of the limitations in image quality implied by atmospheric turbulence. However, MUSE will soon be equipped with an adaptive optics module, with which the achievable spatial resolution could decrease to 0\farc{3} over the $1'\times1'$ field of view. This brings similar studies {(e.g., Izzo et al., in preparation)} into the realm of possibility for at least a handful of close GRBs and will allow us to reassess our constraints on GRB progenitors on a statistically more significant sample.

\begin{acknowledgements}

{We are grateful to the referee as well as S. Schulze, M. {Micha{\l}owski}, G. Leloudas, and D.~A.~Kann for very helpful and constructive comments that increased the quality and strength of this paper.} It is a pleasure to thank D.~Malesani for providing broadband imaging of the SN field and L.~Christensen, J.~Greiner, R.~Yates, T.W.~Chen, J.~Graham, and P.~Wiseman for helpful discussions. T.K. and P.S. acknowledge support through the Sofja Kovalevskaja Award to Patricia Schady from the Alexander von Humboldt Foundation of Germany. L.G. was supported in part by the US National Science Foundation under Grant AST-1311862. We acknowledge the use of \texttt{NumPy} and \texttt{SciPy} \citep{Walt:2011:NAS:1957373.1957466} for computing and \texttt{matplotlib} \citep{Hunter:2007} for creating all plots in this manuscript. 

\end{acknowledgements}

\bibliography{./bibtex/refs}

\begin{thebibliography}{131}
\expandafter\ifx\csname natexlab\endcsname\relax\def\natexlab#1{#1}\fi

\bibitem[{{Alloin} {et~al.}(1979){Alloin}, {Collin-Souffrin}, {Joly}, \&
  {Vigroux}}]{1979A&A....78..200A}
{Alloin}, D., {Collin-Souffrin}, S., {Joly}, M., \& {Vigroux}, L. 1979, \aap,
  78, 200

\bibitem[{{Amor{\'{\i}}n} {et~al.}(2010){Amor{\'{\i}}n}, {P{\'e}rez-Montero},
  \& {V{\'{\i}}lchez}}]{2010ApJ...715L.128A}
{Amor{\'{\i}}n}, R.~O., {P{\'e}rez-Montero}, E., \& {V{\'{\i}}lchez}, J.~M.
  2010, \apjl, 715, L128

\bibitem[{{Anderson} {et~al.}(2010){Anderson}, {Covarrubias}, {James}, {Hamuy},
  \& {Habergham}}]{2010MNRAS.407.2660A}
{Anderson}, J.~P., {Covarrubias}, R.~A., {James}, P.~A., {Hamuy}, M., \&
  {Habergham}, S.~M. 2010, \mnras, 407, 2660

\bibitem[{{Appenzeller} {et~al.}(1998){Appenzeller}, {Fricke}, {F{\"u}rtig},
  {G{\"a}ssler}, {H{\"a}fner}, {Harke}, {Hess}, {Hummel}, {J{\"u}rgens},
  {Kudritzki}, {Mantel}, {Meisl}, {Muschielok}, {Nicklas}, {Rupprecht},
  {Seifert}, {Stahl}, {Szeifert}, \& {Tarantik}}]{1998Msngr..94....1A}
{Appenzeller}, I., {Fricke}, K., {F{\"u}rtig}, W., {et~al.} 1998, The
  Messenger, 94, 1

\bibitem[{{Arabsalmani} {et~al.}(2015){Arabsalmani}, {Roychowdhury}, {Zwaan},
  {Kanekar}, \& {Micha{\l}owski}}]{2015MNRAS.454L..51A}
{Arabsalmani}, M., {Roychowdhury}, S., {Zwaan}, M.~A., {Kanekar}, N., \&
  {Micha{\l}owski}, M.~J. 2015, \mnras, 454, L51

\bibitem[{{Asplund} {et~al.}(2009){Asplund}, {Grevesse}, {Sauval}, \&
  {Scott}}]{2009ARA&A..47..481A}
{Asplund}, M., {Grevesse}, N., {Sauval}, A.~J., \& {Scott}, P. 2009, \araa, 47,
  481

\bibitem[{{Bacon} {et~al.}(2010){Bacon}, {Accardo}, {Adjali}, {Anwand},
  {Bauer}, {Biswas}, {Blaizot}, {Boudon}, {Brau-Nogue}, {Brinchmann},
  {Caillier}, {Capoani}, {Carollo}, {Contini}, {Couderc}, {Daguis{\'e}},
  {Deiries}, {Delabre}, {Dreizler}, {Dubois}, {Dupieux}, {Dupuy}, {Emsellem},
  {Fechner}, {Fleischmann}, {Fran{\c c}ois}, {Gallou}, {Gharsa}, {Glindemann},
  {Gojak}, {Guiderdoni}, {Hansali}, {Hahn}, {Jarno}, {Kelz}, {Koehler},
  {Kosmalski}, {Laurent}, {Le Floch}, {Lilly}, {Lizon}, {Loupias}, {Manescau},
  {Monstein}, {Nicklas}, {Olaya}, {Pares}, {Pasquini}, {P{\'e}contal-Rousset},
  {Pell{\'o}}, {Petit}, {Popow}, {Reiss}, {Remillieux}, {Renault}, {Roth},
  {Rupprecht}, {Serre}, {Schaye}, {Soucail}, {Steinmetz}, {Streicher}, {Stuik},
  {Valentin}, {Vernet}, {Weilbacher}, {Wisotzki}, \&
  {Yerle}}]{2010SPIE.7735E..08B}
{Bacon}, R., {Accardo}, M., {Adjali}, L., {et~al.} 2010, in \procspie, Vol.
  7735, Ground-based and Airborne Instrumentation for Astronomy III, 773508

\bibitem[{{Baldwin} {et~al.}(1981){Baldwin}, {Phillips}, \&
  {Terlevich}}]{1981PASP...93....5B}
{Baldwin}, J.~A., {Phillips}, M.~M., \& {Terlevich}, R. 1981, \pasp, 93, 5

\bibitem[{{Banerjee} {et~al.}(2012){Banerjee}, {Kroupa}, \&
  {Oh}}]{2012ApJ...746...15B}
{Banerjee}, S., {Kroupa}, P., \& {Oh}, S. 2012, \apj, 746, 15

\bibitem[{{Bian} {et~al.}(2016){Bian}, {Kewley}, {Dopita}, \&
  {Blanc}}]{2016arXiv161108595B}
{Bian}, F., {Kewley}, L., {Dopita}, M., \& {Blanc}, G. 2016, \apj, accepted,
  arXiv:1611.08595 [\eprint[arXiv]{1611.08595}]

\bibitem[{{Binette} {et~al.}(2012){Binette}, {Matadamas}, {H{\"a}gele},
  {Nicholls}, {Magris C.}, {Pe{\~n}a-Guerrero}, {Morisset}, \&
  {Rodr{\'{\i}}guez-Gonz{\'a}lez}}]{2012A&A...547A..29B}
{Binette}, L., {Matadamas}, R., {H{\"a}gele}, G.~F., {et~al.} 2012, \aap, 547,
  A29

\bibitem[{{Blanchard} {et~al.}(2016){Blanchard}, {Berger}, \&
  {Fong}}]{2016ApJ...817..144B}
{Blanchard}, P.~K., {Berger}, E., \& {Fong}, W.-f. 2016, \apj, 817, 144

\bibitem[{{Brinchmann} {et~al.}(2008){Brinchmann}, {Pettini}, \&
  {Charlot}}]{2008MNRAS.385..769B}
{Brinchmann}, J., {Pettini}, M., \& {Charlot}, S. 2008, \mnras, 385, 769

\bibitem[{{Bruzual} \& {Charlot}(2003)}]{2003MNRAS.344.1000B}
{Bruzual}, G. \& {Charlot}, S. 2003, \mnras, 344, 1000

\bibitem[{{Cano} {et~al.}(2016){Cano}, {Wang}, {Dai}, \&
  {Wu}}]{2016LPICo1962.4116C}
{Cano}, Z., {Wang}, S.-Q., {Dai}, Z.-G., \& {Wu}, X.-F. 2016, LPI
  Contributions, 1962, 4116

\bibitem[{{Chabrier}(2003)}]{2003PASP..115..763C}
{Chabrier}, G. 2003, \pasp, 115, 763

\bibitem[{{Chen} {et~al.}(2013){Chen}, {Smartt}, {Bresolin}, {Pastorello},
  {Kudritzki}, {Kotak}, {McCrum}, {Fraser}, \& {Valenti}}]{2013ApJ...763L..28C}
{Chen}, T.-W., {Smartt}, S.~J., {Bresolin}, F., {et~al.} 2013, \apjl, 763, L28

\bibitem[{{Christensen} {et~al.}(2008){Christensen}, {Vreeswijk}, {Sollerman},
  {Th{\"o}ne}, {Le Floc'h}, \& {Wiersema}}]{2008A&A...490...45C}
{Christensen}, L., {Vreeswijk}, P.~M., {Sollerman}, J., {et~al.} 2008, \aap,
  490, 45

\bibitem[{{Cid Fernandes} {et~al.}(2005){Cid Fernandes}, {Mateus}, {Sodr{\'e}},
  {Stasi{\'n}ska}, \& {Gomes}}]{2005MNRAS.358..363C}
{Cid Fernandes}, R., {Mateus}, A., {Sodr{\'e}}, L., {Stasi{\'n}ska}, G., \&
  {Gomes}, J.~M. 2005, \mnras, 358, 363

\bibitem[{{Cid Fernandes} {et~al.}(2009){Cid Fernandes}, {Schoenell}, {Gomes},
  {Asari}, {Schlickmann}, {Mateus}, {Stasinska}, {Sodr{\'e}}, {Torres-Papaqui},
  \& {Seagal Collaboration}}]{2009RMxAC..35..127C}
{Cid Fernandes}, R., {Schoenell}, W., {Gomes}, J.~M., {et~al.} 2009, in Rev.
  Mex. Astron. Astrofis. Conf. Ser., Vol.~35, 127--132

\bibitem[{{Clocchiatti} {et~al.}(2011){Clocchiatti}, {Suntzeff}, {Covarrubias},
  \& {Candia}}]{2011AJ....141..163C}
{Clocchiatti}, A., {Suntzeff}, N.~B., {Covarrubias}, R., \& {Candia}, P. 2011,
  \aj, 141, 163

\bibitem[{{Diaz} {et~al.}(1991){Diaz}, {Terlevich}, {Vilchez}, {Pagel}, \&
  {Edmunds}}]{1991MNRAS.253..245D}
{Diaz}, A.~I., {Terlevich}, E., {Vilchez}, J.~M., {Pagel}, B.~E.~J., \&
  {Edmunds}, M.~G. 1991, \mnras, 253, 245

\bibitem[{{Dopita} {et~al.}(2000){Dopita}, {Kewley}, {Heisler}, \&
  {Sutherland}}]{2000ApJ...542..224D}
{Dopita}, M.~A., {Kewley}, L.~J., {Heisler}, C.~A., \& {Sutherland}, R.~S.
  2000, \apj, 542, 224

\bibitem[{{Dopita} {et~al.}(2016){Dopita}, {Kewley}, {Sutherland}, \&
  {Nicholls}}]{2016Ap&SS.361...61D}
{Dopita}, M.~A., {Kewley}, L.~J., {Sutherland}, R.~S., \& {Nicholls}, D.~C.
  2016, \apss, 361, 61

\bibitem[{{Dors} {et~al.}(2011){Dors}, {Krabbe}, {H{\"a}gele}, \&
  {P{\'e}rez-Montero}}]{2011MNRAS.415.3616D}
{Dors}, Jr., O.~L., {Krabbe}, A., {H{\"a}gele}, G.~F., \& {P{\'e}rez-Montero},
  E. 2011, \mnras, 415, 3616

\bibitem[{{Eldridge} {et~al.}(2011){Eldridge}, {Langer}, \&
  {Tout}}]{2011MNRAS.414.3501E}
{Eldridge}, J.~J., {Langer}, N., \& {Tout}, C.~A. 2011, \mnras, 414, 3501

\bibitem[{{Erb} {et~al.}(2006){Erb}, {Shapley}, {Pettini}, {Steidel}, {Reddy},
  \& {Adelberger}}]{2006ApJ...644..813E}
{Erb}, D.~K., {Shapley}, A.~E., {Pettini}, M., {et~al.} 2006, \apj, 644, 813

\bibitem[{{Esteban} {et~al.}(2004){Esteban}, {Peimbert}, {Garc{\'{\i}}a-Rojas},
  {Ruiz}, {Peimbert}, \& {Rodr{\'{\i}}guez}}]{2004MNRAS.355..229E}
{Esteban}, C., {Peimbert}, M., {Garc{\'{\i}}a-Rojas}, J., {et~al.} 2004,
  \mnras, 355, 229

\bibitem[{{Evans} \& {Dopita}(1985)}]{1985ApJS...58..125E}
{Evans}, I.~N. \& {Dopita}, M.~A. 1985, \apjs, 58, 125

\bibitem[{{Fagotto} {et~al.}(1994){Fagotto}, {Bressan}, {Bertelli}, \&
  {Chiosi}}]{1994A&AS..105...29F}
{Fagotto}, F., {Bressan}, A., {Bertelli}, G., \& {Chiosi}, C. 1994, \aaps, 105

\bibitem[{{F{\"o}rster Schreiber} {et~al.}(2009){F{\"o}rster Schreiber},
  {Genzel}, {Bouch{\'e}}, {Cresci}, {Davies}, {Buschkamp}, {Shapiro},
  {Tacconi}, {Hicks}, {Genel}, {Shapley}, {Erb}, {Steidel}, {Lutz},
  {Eisenhauer}, {Gillessen}, {Sternberg}, {Renzini}, {Cimatti}, {Daddi},
  {Kurk}, {Lilly}, {Kong}, {Lehnert}, {Nesvadba}, {Verma}, {McCracken},
  {Arimoto}, {Mignoli}, \& {Onodera}}]{2009ApJ...706.1364F}
{F{\"o}rster Schreiber}, N.~M., {Genzel}, R., {Bouch{\'e}}, N., {et~al.} 2009,
  \apj, 706, 1364

\bibitem[{{Fruchter} {et~al.}(2006){Fruchter}, {Levan}, {Strolger},
  {Vreeswijk}, {Thorsett}, {Bersier}, {Burud}, {Castro Cer{\'o}n},
  {Castro-Tirado}, {Conselice}, {Dahlen}, {Ferguson}, {Fynbo}, {Garnavich},
  {Gibbons}, {Gorosabel}, {Gull}, {Hjorth}, {Holland}, {Kouveliotou}, {Levay},
  {Livio}, {Metzger}, {Nugent}, {Petro}, {Pian}, {Rhoads}, {Riess}, {Sahu},
  {Smette}, {Tanvir}, {Wijers}, \& {Woosley}}]{2006Natur.441..463F}
{Fruchter}, A.~S., {Levan}, A.~J., {Strolger}, L., {et~al.} 2006, \nat, 441,
  463

\bibitem[{{Fynbo} {et~al.}(2009){Fynbo}, {Jakobsson}, {Prochaska}, {Malesani},
  {Ledoux}, {de Ugarte Postigo}, {Nardini}, {Vreeswijk}, {Wiersema}, {Hjorth},
  {Sollerman}, {Chen}, {Th{\"o}ne}, {Bj{\"o}rnsson}, {Bloom}, {Castro-Tirado},
  {Christensen}, {De Cia}, {Fruchter}, {Gorosabel}, {Graham}, {Jaunsen},
  {Jensen}, {Kann}, {Kouveliotou}, {Levan}, {Maund}, {Masetti},
  {Milvang-Jensen}, {Palazzi}, {Perley}, {Pian}, {Rol}, {Schady}, {Starling},
  {Tanvir}, {Watson}, {Xu}, {Augusteijn}, {Grundahl}, {Telting}, \&
  {Quirion}}]{2009ApJS..185..526F}
{Fynbo}, J.~P.~U., {Jakobsson}, P., {Prochaska}, J.~X., {et~al.} 2009, \apjs,
  185, 526

\bibitem[{{Fynbo} {et~al.}(2000){Fynbo}, {Holland}, {Andersen}, {Thomsen},
  {Hjorth}, {Bj{\"o}rnsson}, {Jaunsen}, {Natarajan}, \&
  {Tanvir}}]{2000ApJ...542L..89F}
{Fynbo}, J.~U., {Holland}, S., {Andersen}, M.~I., {et~al.} 2000, \apjl, 542,
  L89

\bibitem[{{Galama} {et~al.}(1998){Galama}, {Vreeswijk}, {van Paradijs},
  {Kouveliotou}, {Augusteijn}, {B{\"o}hnhardt}, {Brewer}, {Doublier},
  {Gonzalez}, {Leibundgut}, {Lidman}, {Hainaut}, {Patat}, {Heise}, {in't Zand},
  {Hurley}, {Groot}, {Strom}, {Mazzali}, {Iwamoto}, {Nomoto}, {Umeda},
  {Nakamura}, {Young}, {Suzuki}, {Shigeyama}, {Koshut}, {Kippen}, {Robinson},
  {de Wildt}, {Wijers}, {Tanvir}, {Greiner}, {Pian}, {Palazzi}, {Frontera},
  {Masetti}, {Nicastro}, {Feroci}, {Costa}, {Piro}, {Peterson}, {Tinney},
  {Boyle}, {Cannon}, {Stathakis}, {Sadler}, {Begam}, \&
  {Ianna}}]{1998Natur.395..670G}
{Galama}, T.~J., {Vreeswijk}, P.~M., {van Paradijs}, J., {et~al.} 1998, \nat,
  395, 670

\bibitem[{{Galbany} {et~al.}(2016{\natexlab{a}}){Galbany}, {Anderson},
  {Rosales-Ortega}, {Kuncarayakti}, {Kr{\"u}hler}, {S{\'a}nchez},
  {Falc{\'o}n-Barroso}, {P{\'e}rez}, {Maureira}, {Hamuy},
  {Gonz{\'a}lez-Gait{\'a}n}, {F{\"o}rster}, \& {Moral}}]{2016MNRAS.455.4087G}
{Galbany}, L., {Anderson}, J.~P., {Rosales-Ortega}, F.~F., {et~al.}
  2016{\natexlab{a}}, \mnras, 455, 4087

\bibitem[{{Galbany} {et~al.}(2014){Galbany}, {Stanishev}, {Mour{\~a}o},
  {Rodrigues}, {Flores}, {Garc{\'{\i}}a-Benito}, {Mast}, {Mendoza},
  {S{\'a}nchez}, {Badenes}, {Barrera-Ballesteros}, {Bland-Hawthorn},
  {Falc{\'o}n-Barroso}, {Garc{\'{\i}}a-Lorenzo}, {Gomes}, {Gonz{\'a}lez
  Delgado}, {Kehrig}, {Lyubenova}, {L{\'o}pez-S{\'a}nchez}, {de
  Lorenzo-C{\'a}ceres}, {Marino}, {Meidt}, {Moll{\'a}}, {Papaderos},
  {P{\'e}rez-Torres}, {Rosales-Ortega}, \& {van de Ven}}]{2014A&A...572A..38G}
{Galbany}, L., {Stanishev}, V., {Mour{\~a}o}, A.~M., {et~al.} 2014, \aap, 572,
  A38

\bibitem[{{Galbany} {et~al.}(2016{\natexlab{b}}){Galbany}, {Stanishev},
  {Mour{\~a}o}, {Rodrigues}, {Flores}, {Walcher}, {S{\'a}nchez},
  {Garc{\'{\i}}a-Benito}, {Mast}, {Badenes}, {Gonz{\'a}lez Delgado}, {Kehrig},
  {Lyubenova}, {Marino}, {Moll{\'a}}, {Meidt}, {P{\'e}rez}, {van de Ven}, \&
  {V{\'{\i}}lchez}}]{2016A&A...591A..48G}
{Galbany}, L., {Stanishev}, V., {Mour{\~a}o}, A.~M., {et~al.}
  2016{\natexlab{b}}, \aap, 591, A48

\bibitem[{{Gonz{\'a}lez Delgado} {et~al.}(1999){Gonz{\'a}lez Delgado},
  {Leitherer}, \& {Heckman}}]{1999ApJS..125..489G}
{Gonz{\'a}lez Delgado}, R.~M., {Leitherer}, C., \& {Heckman}, T.~M. 1999,
  \apjs, 125, 489

\bibitem[{{Graham} \& {Fruchter}(2013)}]{2013ApJ...774..119G}
{Graham}, J.~F. \& {Fruchter}, A.~S. 2013, \apj, 774, 119

\bibitem[{{Hammer} {et~al.}(2006){Hammer}, {Flores}, {Schaerer},
  {Dessauges-Zavadsky}, {Le Floc'h}, \& {Puech}}]{2006A&A...454..103H}
{Hammer}, F., {Flores}, H., {Schaerer}, D., {et~al.} 2006, \aap, 454, 103

\bibitem[{{Hjorth} \& {Bloom}(2012)}]{2012grbu.book..169H}
{Hjorth}, J. \& {Bloom}, J.~S. 2012, {The Gamma-Ray Burst - Supernova
  Connection}, 169--190

\bibitem[{{Ho} {et~al.}(2015){Ho}, {Kudritzki}, {Kewley}, {Zahid}, {Dopita},
  {Bresolin}, \& {Rupke}}]{2015MNRAS.448.2030H}
{Ho}, I.-T., {Kudritzki}, R.-P., {Kewley}, L.~J., {et~al.} 2015, \mnras, 448,
  2030

\bibitem[{{Hoogerwerf} {et~al.}(2001){Hoogerwerf}, {de Bruijne}, \& {de
  Zeeuw}}]{2001A&A...365...49H}
{Hoogerwerf}, R., {de Bruijne}, J.~H.~J., \& {de Zeeuw}, P.~T. 2001, \aap, 365,
  49

\bibitem[{Hunter(2007)}]{Hunter:2007}
Hunter, J.~D. 2007, Computing In Science \& Engineering, 9, 90

\bibitem[{{Iwamoto} {et~al.}(1998){Iwamoto}, {Mazzali}, {Nomoto}, {Umeda},
  {Nakamura}, {Patat}, {Danziger}, {Young}, {Suzuki}, {Shigeyama},
  {Augusteijn}, {Doublier}, {Gonzalez}, {Boehnhardt}, {Brewer}, {Hainaut},
  {Lidman}, {Leibundgut}, {Cappellaro}, {Turatto}, {Galama}, {Vreeswijk},
  {Kouveliotou}, {van Paradijs}, {Pian}, {Palazzi}, \&
  {Frontera}}]{1998Natur.395..672I}
{Iwamoto}, K., {Mazzali}, P.~A., {Nomoto}, K., {et~al.} 1998, \nat, 395, 672

\bibitem[{{Izotov} {et~al.}(2006){Izotov}, {Stasi{\'n}ska}, {Meynet}, {Guseva},
  \& {Thuan}}]{2006A&A...448..955I}
{Izotov}, Y.~I., {Stasi{\'n}ska}, G., {Meynet}, G., {Guseva}, N.~G., \&
  {Thuan}, T.~X. 2006, \aap, 448, 955

\bibitem[{{Izotov} \& {Thuan}(1999)}]{1999ApJ...511..639I}
{Izotov}, Y.~I. \& {Thuan}, T.~X. 1999, \apj, 511, 639

\bibitem[{{Japelj} {et~al.}(2016){Japelj}, {Vergani}, {Salvaterra}, {D'Avanzo},
  {Mannucci}, {Fernandez-Soto}, {Boissier}, {Hunt}, {Atek},
  {Rodr{\'{\i}}guez-Mu{\~n}oz}, {Scodeggio}, {Cristiani}, {Le Floc'h},
  {Flores}, {Gallego}, {Ghirlanda}, {Gomboc}, {Hammer}, {Perley}, {Pescalli},
  {Petitjean}, {Puech}, {Rafelski}, \& {Tagliaferri}}]{2016A&A...590A.129J}
{Japelj}, J., {Vergani}, S.~D., {Salvaterra}, R., {et~al.} 2016, \aap, 590,
  A129

\bibitem[{{Kann} {et~al.}(2016){Kann}, {Schady}, {Olivares E.}, {Klose},
  {Rossi}, {Perley}, {Kr{\"u}hler}, {Greiner}, {Nicuesa Guelbenzu}, {Elliott},
  {Knust}, {Filgas}, {Pian}, {Mazzali}, {Fynbo}, {Leloudas}, {Afonso},
  {Delvaux}, {Graham}, {Rau}, {Schmidl}, {Schulze}, {Tanga}, {Updike}, \&
  {Varela}}]{2016arXiv160606791K}
{Kann}, D.~A., {Schady}, P., {Olivares E.}, F., {et~al.} 2016, A\&A, submitted
  [\eprint[arXiv]{1606.06791}]

\bibitem[{{Kashino} {et~al.}(2016){Kashino}, {Renzini}, {Silverman}, \&
  {Daddi}}]{2016ApJ...823L..24K}
{Kashino}, D., {Renzini}, A., {Silverman}, J.~D., \& {Daddi}, E. 2016, \apjl,
  823, L24

\bibitem[{{Kennicutt}(1998)}]{1998ARA&A..36..189K}
{Kennicutt}, Jr., R.~C. 1998, \araa, 36, 189

\bibitem[{{Kewley} \& {Dopita}(2002)}]{2002ApJS..142...35K}
{Kewley}, L.~J. \& {Dopita}, M.~A. 2002, \apjs, 142, 35

\bibitem[{{Kewley} {et~al.}(2013){Kewley}, {Dopita}, {Leitherer}, {Dav{\'e}},
  {Yuan}, {Allen}, {Groves}, \& {Sutherland}}]{2013ApJ...774..100K}
{Kewley}, L.~J., {Dopita}, M.~A., {Leitherer}, C., {et~al.} 2013, \apj, 774,
  100

\bibitem[{{Kewley} \& {Ellison}(2008)}]{2008ApJ...681.1183K}
{Kewley}, L.~J. \& {Ellison}, S.~L. 2008, \apj, 681, 1183

\bibitem[{{Kobulnicky} \& {Kewley}(2004)}]{2004ApJ...617..240K}
{Kobulnicky}, H.~A. \& {Kewley}, L.~J. 2004, \apj, 617, 240

\bibitem[{{Kr{\"u}hler} {et~al.}(2012{\natexlab{a}}){Kr{\"u}hler}, {Fynbo},
  {Geier}, {Hjorth}, {Malesani}, {Milvang-Jensen}, {Levan}, {Sparre}, {Watson},
  \& {Zafar}}]{2012A&A...546A...8K}
{Kr{\"u}hler}, T., {Fynbo}, J.~P.~U., {Geier}, S., {et~al.} 2012{\natexlab{a}},
  \aap, 546, A8

\bibitem[{{Kr{\"u}hler} {et~al.}(2015){Kr{\"u}hler}, {Malesani}, {Fynbo},
  {Hartoog}, {Hjorth}, {Jakobsson}, {Perley}, {Rossi}, {Schady}, {Schulze},
  {Tanvir}, {Vergani}, {Wiersema}, {Afonso}, {Bolmer}, {Cano}, {Covino},
  {D'Elia}, {de Ugarte Postigo}, {Filgas}, {Friis}, {Graham}, {Greiner},
  {Goldoni}, {Gomboc}, {Hammer}, {Japelj}, {Kann}, {Kaper}, {Klose}, {Levan},
  {Leloudas}, {Milvang-Jensen}, {Nicuesa Guelbenzu}, {Palazzi}, {Pian},
  {Piranomonte}, {S{\'a}nchez-Ram{\'{\i}}rez}, {Savaglio}, {Selsing},
  {Tagliaferri}, {Vreeswijk}, {Watson}, \& {Xu}}]{2015A&A...581A.125K}
{Kr{\"u}hler}, T., {Malesani}, D., {Fynbo}, J.~P.~U., {et~al.} 2015, \aap, 581,
  A125

\bibitem[{{Kr{\"u}hler} {et~al.}(2012{\natexlab{b}}){Kr{\"u}hler}, {Malesani},
  {Milvang-Jensen}, {Fynbo}, {Hjorth}, {Jakobsson}, {Levan}, {Sparre},
  {Tanvir}, \& {Watson}}]{2012ApJ...758...46K}
{Kr{\"u}hler}, T., {Malesani}, D., {Milvang-Jensen}, B., {et~al.}
  2012{\natexlab{b}}, \apj, 758, 46

\bibitem[{{Kulkarni} {et~al.}(1998){Kulkarni}, {Frail}, {Wieringa}, {Ekers},
  {Sadler}, {Wark}, {Higdon}, {Phinney}, \& {Bloom}}]{1998Natur.395..663K}
{Kulkarni}, S.~R., {Frail}, D.~A., {Wieringa}, M.~H., {et~al.} 1998, \nat, 395,
  663

\bibitem[{{Kuncarayakti} {et~al.}(2013{\natexlab{a}}){Kuncarayakti}, {Doi},
  {Aldering}, {Arimoto}, {Maeda}, {Morokuma}, {Pereira}, {Usuda}, \&
  {Hashiba}}]{2013AJ....146...30K}
{Kuncarayakti}, H., {Doi}, M., {Aldering}, G., {et~al.} 2013{\natexlab{a}},
  \aj, 146, 30

\bibitem[{{Kuncarayakti} {et~al.}(2013{\natexlab{b}}){Kuncarayakti}, {Doi},
  {Aldering}, {Arimoto}, {Maeda}, {Morokuma}, {Pereira}, {Usuda}, \&
  {Hashiba}}]{2013AJ....146...31K}
{Kuncarayakti}, H., {Doi}, M., {Aldering}, G., {et~al.} 2013{\natexlab{b}},
  \aj, 146, 31

\bibitem[{{Kuncarayakti} {et~al.}(2016){Kuncarayakti}, {Galbany}, {Anderson},
  {Kr{\"u}hler}, \& {Hamuy}}]{2016arXiv160703446K}
{Kuncarayakti}, H., {Galbany}, L., {Anderson}, J.~P., {Kr{\"u}hler}, T., \&
  {Hamuy}, M. 2016, \aap, 593, A78

\bibitem[{{Lauberts} \& {Valentijn}(1989)}]{1989spce.book.....L}
{Lauberts}, A. \& {Valentijn}, E.~A. 1989, {The surface photometry catalogue of
  the ESO-Uppsala galaxies}

\bibitem[{{Le Floc'h} {et~al.}(2012){Le Floc'h}, {Charmandaris}, {Gordon},
  {Forrest}, {Brandl}, {Schaerer}, {Dessauges-Zavadsky}, \&
  {Armus}}]{2012ApJ...746....7L}
{Le Floc'h}, E., {Charmandaris}, V., {Gordon}, K., {et~al.} 2012, \apj, 746, 7

\bibitem[{{Leitherer} {et~al.}(1999){Leitherer}, {Schaerer}, {Goldader},
  {Delgado}, {Robert}, {Kune}, {de Mello}, {Devost}, \&
  {Heckman}}]{1999ApJS..123....3L}
{Leitherer}, C., {Schaerer}, D., {Goldader}, J.~D., {et~al.} 1999, \apjs, 123,
  3

\bibitem[{{Leloudas} {et~al.}(2011){Leloudas}, {Gallazzi}, {Sollerman},
  {Stritzinger}, {Fynbo}, {Hjorth}, {Malesani}, {Micha{\l}owski},
  {Milvang-Jensen}, \& {Smith}}]{2011A&A...530A..95L}
{Leloudas}, G., {Gallazzi}, A., {Sollerman}, J., {et~al.} 2011, \aap, 530, A95

\bibitem[{{Leloudas} {et~al.}(2015){Leloudas}, {Schulze}, {Kr{\"u}hler},
  {Gorosabel}, {Christensen}, {Mehner}, {de Ugarte Postigo}, {Amor{\'{\i}}n},
  {Th{\"o}ne}, {Anderson}, {Bauer}, {Gallazzi}, {He{\l}miniak}, {Hjorth},
  {Ibar}, {Malesani}, {Morell}, {Vinko}, \& {Wheeler}}]{2014arXiv1409.8331L}
{Leloudas}, G., {Schulze}, S., {Kr{\"u}hler}, T., {et~al.} 2015, \mnras, 449,
  917

\bibitem[{{Levesque} {et~al.}(2011){Levesque}, {Berger}, {Soderberg}, \&
  {Chornock}}]{2011ApJ...739...23L}
{Levesque}, E.~M., {Berger}, E., {Soderberg}, A.~M., \& {Chornock}, R. 2011,
  \apj, 739, 23

\bibitem[{{Levesque} \& {Leitherer}(2013)}]{2013ApJ...779..170L}
{Levesque}, E.~M. \& {Leitherer}, C. 2013, \apj, 779, 170

\bibitem[{{Li} {et~al.}(2011){Li}, {Leaman}, {Chornock}, {Filippenko},
  {Poznanski}, {Ganeshalingam}, {Wang}, {Modjaz}, {Jha}, {Foley}, \&
  {Smith}}]{2011MNRAS.412.1441L}
{Li}, W., {Leaman}, J., {Chornock}, R., {et~al.} 2011, \mnras, 412, 1441

\bibitem[{{L{\'o}pez-S{\'a}nchez} {et~al.}(2012){L{\'o}pez-S{\'a}nchez},
  {Dopita}, {Kewley}, {Zahid}, {Nicholls}, \&
  {Scharw{\"a}chter}}]{2012MNRAS.426.2630L}
{L{\'o}pez-S{\'a}nchez}, {\'A}.~R., {Dopita}, M.~A., {Kewley}, L.~J., {et~al.}
  2012, \mnras, 426, 2630

\bibitem[{{Lunnan} {et~al.}(2014){Lunnan}, {Chornock}, {Berger}, {Laskar},
  {Fong}, {Rest}, {Sanders}, {Challis}, {Drout}, {Foley}, {Huber}, {Kirshner},
  {Leibler}, {Marion}, {McCrum}, {Milisavljevic}, {Narayan}, {Scolnic},
  {Smartt}, {Smith}, {Soderberg}, {Tonry}, {Burgett}, {Chambers}, {Flewelling},
  {Hodapp}, {Kaiser}, {Magnier}, {Price}, \& {Wainscoat}}]{2014ApJ...787..138L}
{Lunnan}, R., {Chornock}, R., {Berger}, E., {et~al.} 2014, \apj, 787, 138

\bibitem[{{Lyman} {et~al.}(2017){Lyman}, {Levan}, {Tanvir}, {Fynbo}, {McGuire},
  {Perley}, {Angus}, {Bloom}, {Conselice}, {Fruchter}, {Hjorth}, {Jakobsson},
  \& {Starling}}]{2017MNRAS.tmp..220L}
{Lyman}, J.~D., {Levan}, A.~J., {Tanvir}, N.~R., {et~al.} 2017, \mnras
  [\eprint[arXiv]{1701.05925}]

\bibitem[{{Maeda} {et~al.}(2006){Maeda}, {Nomoto}, {Mazzali}, \&
  {Deng}}]{2006ApJ...640..854M}
{Maeda}, K., {Nomoto}, K., {Mazzali}, P.~A., \& {Deng}, J. 2006, \apj, 640, 854

\bibitem[{{Maiolino} {et~al.}(2008){Maiolino}, {Nagao}, {Grazian}, {Cocchia},
  {Marconi}, {Mannucci}, {Cimatti}, {Pipino}, {Ballero}, {Calura}, {Chiappini},
  {Fontana}, {Granato}, {Matteucci}, {Pastorini}, {Pentericci}, {Risaliti},
  {Salvati}, \& {Silva}}]{2008A&A...488..463M}
{Maiolino}, R., {Nagao}, T., {Grazian}, A., {et~al.} 2008, \aap, 488, 463

\bibitem[{{Marino} {et~al.}(2013){Marino}, {Rosales-Ortega}, {S{\'a}nchez},
  {Gil de Paz}, {V{\'{\i}}lchez}, {Miralles-Caballero}, {Kehrig},
  {P{\'e}rez-Montero}, {Stanishev}, {Iglesias-P{\'a}ramo}, {D{\'{\i}}az},
  {Castillo-Morales}, {Kennicutt}, {L{\'o}pez-S{\'a}nchez}, {Galbany},
  {Garc{\'{\i}}a-Benito}, {Mast}, {Mendez-Abreu}, {Monreal-Ibero}, {Husemann},
  {Walcher}, {Garc{\'{\i}}a-Lorenzo}, {Masegosa}, {Del Olmo Orozco},
  {Mour{\~a}o}, {Ziegler}, {Moll{\'a}}, {Papaderos},
  {S{\'a}nchez-Bl{\'a}zquez}, {Gonz{\'a}lez Delgado}, {Falc{\'o}n-Barroso},
  {Roth}, {van de Ven}, \& {Califa Team}}]{2013A&A...559A.114M}
{Marino}, R.~A., {Rosales-Ortega}, F.~F., {S{\'a}nchez}, S.~F., {et~al.} 2013,
  \aap, 559, A114

\bibitem[{{Mazzali} {et~al.}(2001){Mazzali}, {Nomoto}, {Patat}, \&
  {Maeda}}]{2001ApJ...559.1047M}
{Mazzali}, P.~A., {Nomoto}, K., {Patat}, F., \& {Maeda}, K. 2001, \apj, 559,
  1047

\bibitem[{{McGaugh}(1991)}]{1991ApJ...380..140M}
{McGaugh}, S.~S. 1991, \apj, 380, 140

\bibitem[{{Mendoza} \& {Zeippen}(1982)}]{1982MNRAS.199.1025M}
{Mendoza}, C. \& {Zeippen}, C.~J. 1982, \mnras, 199, 1025

\bibitem[{{Meynet} \& {Maeder}(2005)}]{2005A&A...429..581M}
{Meynet}, G. \& {Maeder}, A. 2005, \aap, 429, 581

\bibitem[{{Micha{\l}owski} {et~al.}(2016){Micha{\l}owski}, {Castro Ceron},
  {Wardlow}, {Karska}, {Messias}, {van der Werf}, {Hunt}, {Baes},
  {Castro-Tirado}, {Gentile}, {Hjorth}, {Le Floc'h}, {Perez Martinez}, {Nicuesa
  Guelbenzu}, {Rasmussen}, {Rizzo}, {Rossi}, {Sanchez-Portal}, {Schady},
  {Sollerman}, \& {Xu}}]{2016arXiv160901742M}
{Micha{\l}owski}, M.~J., {Castro Ceron}, J.~M., {Wardlow}, J.~L., {et~al.}
  2016, A\&A, in press [\eprint{arXiv:1609.01742}]

\bibitem[{{Micha{\l}owski} {et~al.}(2015){Micha{\l}owski}, {Gentile}, {Hjorth},
  {Krumholz}, {Tanvir}, {Kamphuis}, {Burlon}, {Baes}, {Basa}, {Berta}, {Castro
  Cer{\'o}n}, {Crosby}, {D'Elia}, {Elliott}, {Greiner}, {Hunt}, {Klose},
  {Koprowski}, {Le Floc'h}, {Malesani}, {Murphy}, {Nicuesa Guelbenzu},
  {Palazzi}, {Rasmussen}, {Rossi}, {Savaglio}, {Schady}, {Sollerman}, {de
  Ugarte Postigo}, {Watson}, {van der Werf}, {Vergani}, \&
  {Xu}}]{2015A&A...582A..78M}
{Micha{\l}owski}, M.~J., {Gentile}, G., {Hjorth}, J., {et~al.} 2015, \aap, 582,
  A78

\bibitem[{{Micha{\l}owski} {et~al.}(2009){Micha{\l}owski}, {Hjorth},
  {Malesani}, {Micha{\l}owski}, {Castro Cer{\'o}n}, {Reinfrank}, {Garrett},
  {Fynbo}, {Watson}, \& {J{\o}rgensen}}]{2009ApJ...693..347M}
{Micha{\l}owski}, M.~J., {Hjorth}, J., {Malesani}, D., {et~al.} 2009, \apj,
  693, 347

\bibitem[{{Micha{\l}owski} {et~al.}(2014){Micha{\l}owski}, {Hunt}, {Palazzi},
  {Savaglio}, {Gentile}, {Rasmussen}, {Baes}, {Basa}, {Bianchi}, {Berta},
  {Burlon}, {Castro Cer{\'o}n}, {Covino}, {Cuby}, {D'Elia}, {Ferrero},
  {G{\"o}tz}, {Jhorth}, {Koprowski}, {Le Borgne}, {Le Floc'h}, {Malesani},
  {Murphy}, {Pian}, {Piranomonte}, {Rossi}, {Sollerman}, {Tanvir}, {de Ugarte
  Postigo}, {Watson}, {van der Werf}, {Vergani}, \& {Xu}}]{2014A&A...562A..70M}
{Micha{\l}owski}, M.~J., {Hunt}, L.~K., {Palazzi}, E., {et~al.} 2014, \aap,
  562, A70

\bibitem[{{Modjaz} {et~al.}(2011){Modjaz}, {Kewley}, {Bloom}, {Filippenko},
  {Perley}, \& {Silverman}}]{2011ApJ...731L...4M}
{Modjaz}, M., {Kewley}, L., {Bloom}, J.~S., {et~al.} 2011, \apjl, 731, L4

\bibitem[{{Morisset} {et~al.}(2016){Morisset}, {Delgado-Inglada},
  {S{\'a}nchez}, {Galbany}, {Garc{\'{\i}}a-Benito}, {Husemann}, {Marino},
  {Mast}, \& {Roth}}]{2016A&A...594A..37M}
{Morisset}, C., {Delgado-Inglada}, G., {S{\'a}nchez}, S.~F., {et~al.} 2016,
  \aap, 594, A37

\bibitem[{{Nagao} {et~al.}(2006){Nagao}, {Maiolino}, \&
  {Marconi}}]{2006A&A...459...85N}
{Nagao}, T., {Maiolino}, R., \& {Marconi}, A. 2006, \aap, 459, 85

\bibitem[{{Nicholls} {et~al.}(2012){Nicholls}, {Dopita}, \&
  {Sutherland}}]{2012ApJ...752..148N}
{Nicholls}, D.~C., {Dopita}, M.~A., \& {Sutherland}, R.~S. 2012, \apj, 752, 148

\bibitem[{{Nicholls} {et~al.}(2013){Nicholls}, {Dopita}, {Sutherland},
  {Kewley}, \& {Palay}}]{2013ApJS..207...21N}
{Nicholls}, D.~C., {Dopita}, M.~A., {Sutherland}, R.~S., {Kewley}, L.~J., \&
  {Palay}, E. 2013, \apjs, 207, 21

\bibitem[{{Osterbrock}(1989)}]{1989agna.book.....O}
{Osterbrock}, D.~E. 1989, {Astrophysics of gaseous nebulae and active galactic
  nuclei}

\bibitem[{{Osterbrock} \& {Ferland}(2006)}]{2006agna.book.....O}
{Osterbrock}, D.~E. \& {Ferland}, G.~J. 2006, {Astrophysics of gaseous nebulae
  and active galactic nuclei}

\bibitem[{{Pagel} {et~al.}(1979){Pagel}, {Edmunds}, {Blackwell}, {Chun}, \&
  {Smith}}]{1979MNRAS.189...95P}
{Pagel}, B.~E.~J., {Edmunds}, M.~G., {Blackwell}, D.~E., {Chun}, M.~S., \&
  {Smith}, G. 1979, \mnras, 189, 95

\bibitem[{{Patat} {et~al.}(2001){Patat}, {Cappellaro}, {Danziger}, {Mazzali},
  {Sollerman}, {Augusteijn}, {Brewer}, {Doublier}, {Gonzalez}, {Hainaut},
  {Lidman}, {Leibundgut}, {Nomoto}, {Nakamura}, {Spyromilio}, {Rizzi},
  {Turatto}, {Walsh}, {Galama}, {van Paradijs}, {Kouveliotou}, {Vreeswijk},
  {Frontera}, {Masetti}, {Palazzi}, \& {Pian}}]{2001ApJ...555..900P}
{Patat}, F., {Cappellaro}, E., {Danziger}, J., {et~al.} 2001, \apj, 555, 900

\bibitem[{{Pei}(1992)}]{1992ApJ...395..130P}
{Pei}, Y.~C. 1992, \apj, 395, 130

\bibitem[{{Peimbert}(2003)}]{2003ApJ...584..735P}
{Peimbert}, A. 2003, \apj, 584, 735

\bibitem[{{Peimbert}(1967)}]{1967ApJ...150..825P}
{Peimbert}, M. 1967, \apj, 150, 825

\bibitem[{{Pellegrini} {et~al.}(2011){Pellegrini}, {Baldwin}, \&
  {Ferland}}]{2011ApJ...738...34P}
{Pellegrini}, E.~W., {Baldwin}, J.~A., \& {Ferland}, G.~J. 2011, \apj, 738, 34

\bibitem[{{Perets} \& {{\v S}ubr}(2012)}]{2012ApJ...751..133P}
{Perets}, H.~B. \& {{\v S}ubr}, L. 2012, \apj, 751, 133

\bibitem[{{P{\'e}rez-Montero} {et~al.}(2013){P{\'e}rez-Montero}, {Contini},
  {Lamareille}, {Maier}, {Carollo}, {Kneib}, {Le F{\`e}vre}, {Lilly},
  {Mainieri}, {Renzini}, {Scodeggio}, {Zamorani}, {Bardelli}, {Bolzonella},
  {Bongiorno}, {Caputi}, {Cucciati}, {de la Torre}, {de Ravel}, {Franzetti},
  {Garilli}, {Iovino}, {Kampczyk}, {Knobel}, {Kova{\v c}}, {Le Borgne}, {Le
  Brun}, {Mignoli}, {Pell{\`o}}, {Peng}, {Presotto}, {Ricciardelli},
  {Silverman}, {Tanaka}, {Tasca}, {Tresse}, {Vergani}, \&
  {Zucca}}]{2013A&A...549A..25P}
{P{\'e}rez-Montero}, E., {Contini}, T., {Lamareille}, F., {et~al.} 2013, \aap,
  549, A25

\bibitem[{{P{\'e}rez-Montero} {et~al.}(2016){P{\'e}rez-Montero},
  {Garc{\'{\i}}a-Benito}, {V{\'{\i}}lchez}, {S{\'a}nchez}, {Kehrig},
  {Husemann}, {Duarte Puertas}, {Iglesias-P{\'a}ramo}, {Galbany}, {Moll{\'a}},
  {Walcher}, {Ascas{\'{\i}}bar}, {Gonz{\'a}lez Delgado}, {Marino}, {Masegosa},
  {P{\'e}rez}, {Rosales-Ortega}, {S{\'a}nchez-Bl{\'a}zquez}, {Bland-Hawthorn},
  {Bomans}, {L{\'o}pez-S{\'a}nchez}, {Ziegler}, \& {Califa
  Collaboration}}]{2016A&A...595A..62P}
{P{\'e}rez-Montero}, E., {Garc{\'{\i}}a-Benito}, R., {V{\'{\i}}lchez}, J.~M.,
  {et~al.} 2016, \aap, 595, A62

\bibitem[{{Perley} {et~al.}(2016){Perley}, {Quimby}, {Yan}, {Vreeswijk}, {De
  Cia}, {Lunnan}, {Gal-Yam}, {Yaron}, {Filippenko}, {Graham}, {Laher}, \&
  {Nugent}}]{2016arXiv160408207P}
{Perley}, D.~A., {Quimby}, R.~M., {Yan}, L., {et~al.} 2016, \apj, 830, 13

\bibitem[{{Pettini} \& {Pagel}(2004)}]{2004MNRAS.348L..59P}
{Pettini}, M. \& {Pagel}, B.~E.~J. 2004, \mnras, 348, L59

\bibitem[{{Pilyugin} \& {Thuan}(2005)}]{2005ApJ...631..231P}
{Pilyugin}, L.~S. \& {Thuan}, T.~X. 2005, \apj, 631, 231

\bibitem[{{Planck Collaboration}(2014)}]{2014A&A...571A..16P}
{Planck Collaboration}. 2014, \aap, 571, A16

\bibitem[{{Prieto} {et~al.}(2016){Prieto}, {Kr{\"u}hler}, {Anderson},
  {Galbany}, {Kochanek}, {Aquino}, {Brown}, {Dong}, {F{\"o}rster}, {Holoien},
  {Kuncarayakti}, {Maureira}, {Rosales-Ortega}, {S{\'a}nchez}, {Shappee}, \&
  {Stanek}}]{2016arXiv160900013P}
{Prieto}, J.~L., {Kr{\"u}hler}, T., {Anderson}, J.~P., {et~al.} 2016, \apjl,
  830, L32

\bibitem[{{Prieto} {et~al.}(2008){Prieto}, {Stanek}, \&
  {Beacom}}]{2008ApJ...673..999P}
{Prieto}, J.~L., {Stanek}, K.~Z., \& {Beacom}, J.~F. 2008, \apj, 673, 999

\bibitem[{{S{\'a}nchez} {et~al.}(2014){S{\'a}nchez}, {Rosales-Ortega},
  {Iglesias-P{\'a}ramo}, {Moll{\'a}}, {Barrera-Ballesteros}, {Marino},
  {P{\'e}rez}, {S{\'a}nchez-Blazquez}, {Gonz{\'a}lez Delgado}, {Cid Fernandes},
  {de Lorenzo-C{\'a}ceres}, {Mendez-Abreu}, {Galbany}, {Falcon-Barroso},
  {Miralles-Caballero}, {Husemann}, {Garc{\'{\i}}a-Benito}, {Mast}, {Walcher},
  {Gil de Paz}, {Garc{\'{\i}}a-Lorenzo}, {Jungwiert}, {V{\'{\i}}lchez},
  {J{\'{\i}}lkov{\'a}}, {Lyubenova}, {Cortijo-Ferrero}, {D{\'{\i}}az},
  {Wisotzki}, {M{\'a}rquez}, {Bland-Hawthorn}, {Ellis}, {van de Ven}, {Jahnke},
  {Papaderos}, {Gomes}, {Mendoza}, \&
  {L{\'o}pez-S{\'a}nchez}}]{2014A&A...563A..49S}
{S{\'a}nchez}, S.~F., {Rosales-Ortega}, F.~F., {Iglesias-P{\'a}ramo}, J.,
  {et~al.} 2014, \aap, 563, A49

\bibitem[{{Savaglio} {et~al.}(2009){Savaglio}, {Glazebrook}, \& {Le
  Borgne}}]{2009ApJ...691..182S}
{Savaglio}, S., {Glazebrook}, K., \& {Le Borgne}, D. 2009, \apj, 691, 182

\bibitem[{{Schady} {et~al.}(2015){Schady}, {Kr{\"u}hler}, {Greiner}, {Graham},
  {Kann}, {Bolmer}, {Delvaux}, {Elliott}, {Klose}, {Knust}, {Nicuesa
  Guelbenzu}, {Rau}, {Rossi}, {Savaglio}, {Schmidl}, {Schweyer}, {Sudilovsky},
  {Tanga}, {Tanvir}, {Varela}, \& {Wiseman}}]{2015A&A...579A.126S}
{Schady}, P., {Kr{\"u}hler}, T., {Greiner}, J., {et~al.} 2015, \aap, 579, A126

\bibitem[{{Schlafly} \& {Finkbeiner}(2011)}]{2011ApJ...737..103S}
{Schlafly}, E.~F. \& {Finkbeiner}, D.~P. 2011, \apj, 737, 103

\bibitem[{{Schulze} {et~al.}(2014){Schulze}, {Malesani}, {Cucchiara}, {Tanvir},
  {Kr{\"u}hler}, {de Ugarte Postigo}, {Leloudas}, {Lyman}, {Bersier},
  {Wiersema}, {Perley}, {Schady}, {Gorosabel}, {Anderson}, {Castro-Tirado},
  {Cenko}, {De Cia}, {Ellerbroek}, {Fynbo}, {Greiner}, {Hjorth}, {Kann},
  {Kaper}, {Klose}, {Levan}, {Mart{\'{\i}}n}, {O'Brien}, {Page}, {Pignata},
  {Rapaport}, {S{\'a}nchez-Ram{\'{\i}}rez}, {Sollerman}, {Smith}, {Sparre},
  {Th{\"o}ne}, {Watson}, {Xu}, {Bauer}, {Bayliss}, {Bj{\"o}rnsson}, {Bremer},
  {Cano}, {Covino}, {D'Elia}, {Frail}, {Geier}, {Goldoni}, {Hartoog},
  {Jakobsson}, {Korhonen}, {Lee}, {Milvang-Jensen}, {Nardini}, {Nicuesa
  Guelbenzu}, {Oguri}, {Pandey}, {Petitpas}, {Rossi}, {Sandberg}, {Schmidl},
  {Tagliaferri}, {Tilanus}, {Winters}, {Wright}, \&
  {Wuyts}}]{2014A&A...566A.102S}
{Schulze}, S., {Malesani}, D., {Cucchiara}, A., {et~al.} 2014, \aap, 566, A102

\bibitem[{{Smette} {et~al.}(2015){Smette}, {Sana}, {Noll}, {Horst}, {Kausch},
  {Kimeswenger}, {Barden}, {Szyszka}, {Jones}, {Gallenne}, {Vinther},
  {Ballester}, \& {Taylor}}]{2015A&A...576A..77S}
{Smette}, A., {Sana}, H., {Noll}, S., {et~al.} 2015, \aap, 576, A77

\bibitem[{{Sollerman} {et~al.}(2005){Sollerman}, {{\"O}stlin}, {Fynbo},
  {Hjorth}, {Fruchter}, \& {Pedersen}}]{2005NewA...11..103S}
{Sollerman}, J., {{\"O}stlin}, G., {Fynbo}, J.~P.~U., {et~al.} 2005, \na, 11,
  103

\bibitem[{{Soto} {et~al.}(2016){Soto}, {Lilly}, {Bacon}, {Richard}, \&
  {Conseil}}]{2016MNRAS.458.3210S}
{Soto}, K.~T., {Lilly}, S.~J., {Bacon}, R., {Richard}, J., \& {Conseil}, S.
  2016, \mnras, 458, 3210

\bibitem[{{Stasi{\'n}ska}(2006)}]{2006A&A...454L.127S}
{Stasi{\'n}ska}, G. 2006, \aap, 454, L127

\bibitem[{{Storey} \& {Zeippen}(2000)}]{2000MNRAS.312..813S}
{Storey}, P.~J. \& {Zeippen}, C.~J. 2000, \mnras, 312, 813

\bibitem[{{Svensson} {et~al.}(2010){Svensson}, {Levan}, {Tanvir}, {Fruchter},
  \& {Strolger}}]{2010MNRAS.405...57S}
{Svensson}, K.~M., {Levan}, A.~J., {Tanvir}, N.~R., {Fruchter}, A.~S., \&
  {Strolger}, L.-G. 2010, \mnras, 405, 57

\bibitem[{{Tanvir} {et~al.}(2009){Tanvir}, {Fox}, {Levan}, {Berger},
  {Wiersema}, {Fynbo}, {Cucchiara}, {Kr{\"u}hler}, {Gehrels}, {Bloom},
  {Greiner}, {Evans}, {Rol}, {Olivares}, {Hjorth}, {Jakobsson}, {Farihi},
  {Willingale}, {Starling}, {Cenko}, {Perley}, {Maund}, {Duke}, {Wijers},
  {Adamson}, {Allan}, {Bremer}, {Burrows}, {Castro-Tirado}, {Cavanagh}, {de
  Ugarte Postigo}, {Dopita}, {Fatkhullin}, {Fruchter}, {Foley}, {Gorosabel},
  {Kennea}, {Kerr}, {Klose}, {Krimm}, {Komarova}, {Kulkarni}, {Moskvitin},
  {Mundell}, {Naylor}, {Page}, {Penprase}, {Perri}, {Podsiadlowski}, {Roth},
  {Rutledge}, {Sakamoto}, {Schady}, {Schmidt}, {Soderberg}, {Sollerman},
  {Stephens}, {Stratta}, {Ukwatta}, {Watson}, {Westra}, {Wold}, \&
  {Wolf}}]{2009Natur.461.1254T}
{Tanvir}, N.~R., {Fox}, D.~B., {Levan}, A.~J., {et~al.} 2009, \nat, 461, 1254

\bibitem[{{Th{\"o}ne} {et~al.}(2014){Th{\"o}ne}, {Christensen}, {Prochaska},
  {Bloom}, {Gorosabel}, {Fynbo}, {Jakobsson}, \&
  {Fruchter}}]{2014MNRAS.441.2034T}
{Th{\"o}ne}, C.~C., {Christensen}, L., {Prochaska}, J.~X., {et~al.} 2014,
  \mnras, 441, 2034

\bibitem[{{Th{\"o}ne} {et~al.}(2015){Th{\"o}ne}, {de Ugarte Postigo},
  {Garc{\'{\i}}a-Benito}, {Leloudas}, {Schulze}, \&
  {Amor{\'{\i}}n}}]{2015MNRAS.451L..65T}
{Th{\"o}ne}, C.~C., {de Ugarte Postigo}, A., {Garc{\'{\i}}a-Benito}, R.,
  {et~al.} 2015, \mnras, 451, L65

\bibitem[{{Th{\"o}ne} {et~al.}(2008){Th{\"o}ne}, {Fynbo}, {{\"O}stlin},
  {Milvang-Jensen}, {Wiersema}, {Malesani}, {Ferreira}, {Gorosabel}, {Kann},
  {Watson}, {Micha{\l}owski}, {Fruchter}, {Levan}, {Hjorth}, \&
  {Sollerman}}]{2008ApJ...676.1151T}
{Th{\"o}ne}, C.~C., {Fynbo}, J.~P.~U., {{\"O}stlin}, G., {et~al.} 2008, \apj,
  676, 1151

\bibitem[{{Tremonti} {et~al.}(2004){Tremonti}, {Heckman}, {Kauffmann},
  {Brinchmann}, {Charlot}, {White}, {Seibert}, {Peng}, {Schlegel}, {Uomoto},
  {Fukugita}, \& {Brinkmann}}]{2004ApJ...613..898T}
{Tremonti}, C.~A., {Heckman}, T.~M., {Kauffmann}, G., {et~al.} 2004, \apj, 613,
  898

\bibitem[{{Vergani} {et~al.}(2017){Vergani}, {Palmerio}, {Salvaterra},
  {Japelj}, {Mannucci}, {Perley}, {D'Avanzo}, {Kr{\"u}hler}, {Puech},
  {Boissier}, {Campana}, {Covino}, {Hunt}, {Petitjean}, \&
  {Tagliaferri}}]{2017A&A...599A.120V}
{Vergani}, S.~D., {Palmerio}, J., {Salvaterra}, R., {et~al.} 2017, \aap, 599,
  A120

\bibitem[{Walt {et~al.}(2011)Walt, Colbert, \&
  Varoquaux}]{Walt:2011:NAS:1957373.1957466}
Walt, S. v.~d., Colbert, S.~C., \& Varoquaux, G. 2011, Computing in Science and
  Engg., 13, 22

\bibitem[{{Weilbacher} {et~al.}(2014){Weilbacher}, {Streicher}, {Urrutia},
  {P{\'e}contal-Rousset}, {Jarno}, \& {Bacon}}]{2014ASPC..485..451W}
{Weilbacher}, P.~M., {Streicher}, O., {Urrutia}, T., {et~al.} 2014, in ASP
  Conf. Ser., Vol. 485, Astronomical Data Analysis Software and Systems XXIII,
  ed. N.~{Manset} \& P.~{Forshay}, 451

\bibitem[{{Wesson} {et~al.}(2016){Wesson}, {Stock}, \&
  {Scicluna}}]{2016MNRAS.459.3475W}
{Wesson}, R., {Stock}, D.~J., \& {Scicluna}, P. 2016, \mnras, 459, 3475

\bibitem[{{Wiersema} {et~al.}(2007){Wiersema}, {Savaglio}, {Vreeswijk},
  {Ellison}, {Ledoux}, {Yoon}, {M{\o}ller}, {Sollerman}, {Fynbo}, {Pian},
  {Starling}, \& {Wijers}}]{2007A&A...464..529W}
{Wiersema}, K., {Savaglio}, S., {Vreeswijk}, P.~M., {et~al.} 2007, \aap, 464,
  529

\bibitem[{{Xu} {et~al.}(2013){Xu}, {de Ugarte Postigo}, {Leloudas},
  {Kr{\"u}hler}, {Cano}, {Hjorth}, {Malesani}, {Fynbo}, {Th{\"o}ne},
  {S{\'a}nchez-Ram{\'{\i}}rez}, {Schulze}, {Jakobsson}, {Kaper}, {Sollerman},
  {Watson}, {Cabrera-Lavers}, {Cao}, {Covino}, {Flores}, {Geier}, {Gorosabel},
  {Hu}, {Milvang-Jensen}, {Sparre}, {Xin}, {Zhang}, {Zheng}, \&
  {Zou}}]{2013ApJ...776...98X}
{Xu}, D., {de Ugarte Postigo}, A., {Leloudas}, G., {et~al.} 2013, \apj, 776, 98

\bibitem[{{Zaritsky} {et~al.}(1994){Zaritsky}, {Kennicutt}, \&
  {Huchra}}]{1994ApJ...420...87Z}
{Zaritsky}, D., {Kennicutt}, Jr., R.~C., \& {Huchra}, J.~P. 1994, \apj, 420, 87

\bibitem[{{Zeh} {et~al.}(2004){Zeh}, {Klose}, \&
  {Hartmann}}]{2004ApJ...609..952Z}
{Zeh}, A., {Klose}, S., \& {Hartmann}, D.~H. 2004, \apj, 609, 952

\end{thebibliography}

\begin{appendix}

\section{Analysis of archival long-slit spectra}
\label{app:fors}

The ESO archive contains a number of public, high S/N, but low-resolution spectra obtained with the FOcal Reducer/low dispersion Spectrograph 2 (FORS2; \citealt{1998Msngr..94....1A}), which are particularly interesting in the context of this work. These are the same spectra used by \citet{2006A&A...454..103H} and were taken with a 1\farc{0} slit on 2004-07-15 with the grism 600$B$ (3450~\AA~ to 6050~\AA~ at a resolving power $R$ of $R\sim850$) and on 2004-07-16 with grism 600$RI$ (5300~\AA~ to 8450~\AA~ and $R\sim1050$). The position angle during both observations was 28$^\circ$ from east such that both the SN and WR region are covered by the slit (Fig.~\ref{fig:Host}). The total exposure time was 1350~s (each three individual frames with 300~s and 150~s) in the 600$B$ setup and 2250~s in the 600$RI$ (each five single frames with 300~s and 150~s) setup. We reduced and analyzed the archival FORS2 data using standard procedures, specifically the ESO FORS2 pipeline in its version \texttt{5.3.8} and self-written methods and algorithms in \texttt{python} \citep{2015A&A...581A.125K}.

\begin{figure}
\includegraphics[angle=0, width=0.93\columnwidth]{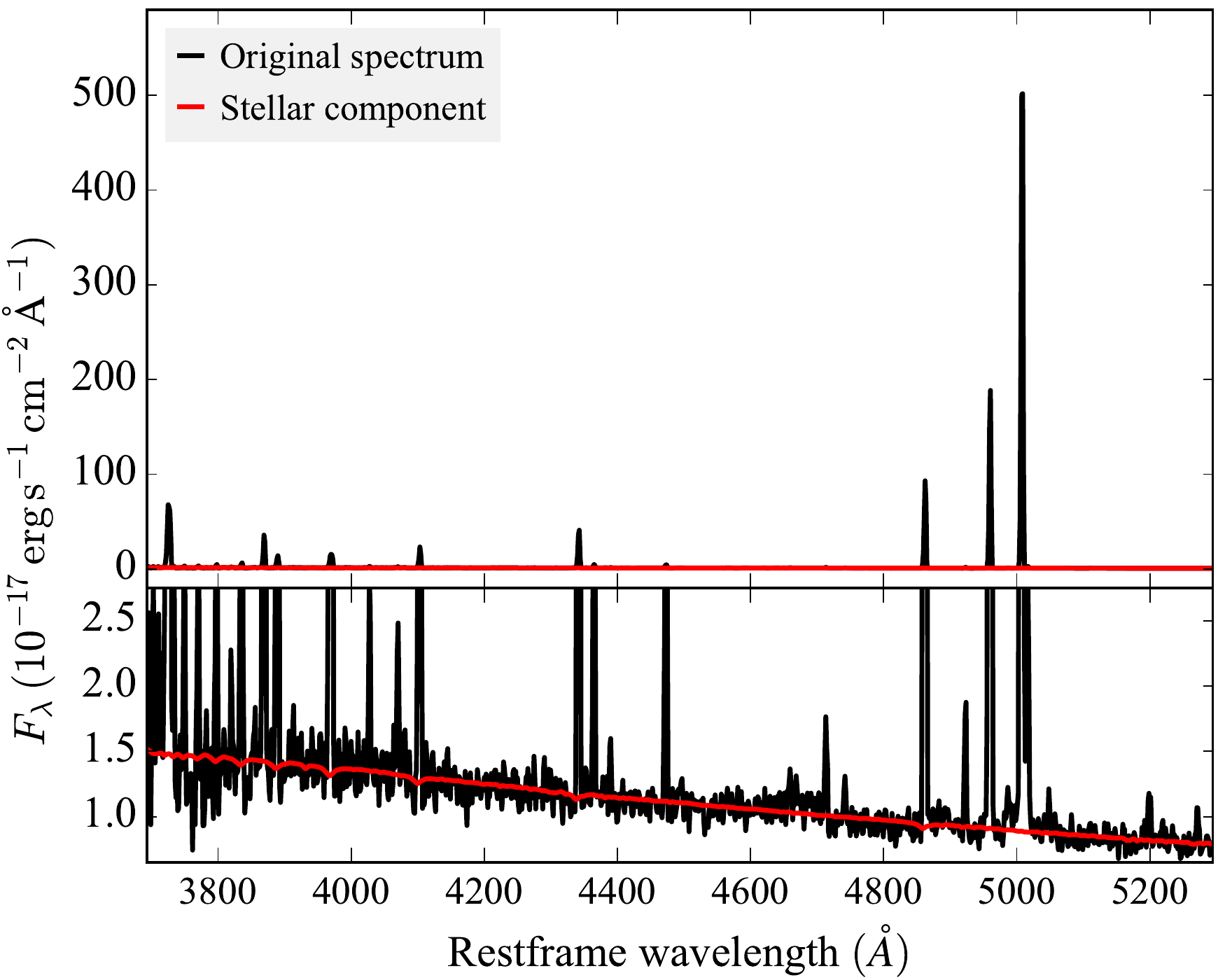}
\caption{FORS2 600$B$ spectrum of the brightest pixel of the WR region. {Black is the original spectrum and red represents the fitted stellar component. The \textit{upper panel} shows the full flux range, while the \textit{lower panel} is a zoom into the stellar continuum.}}
\label{fig:FORSWR}
\end{figure}

\begin{figure}
\includegraphics[angle=0, width=0.93\columnwidth]{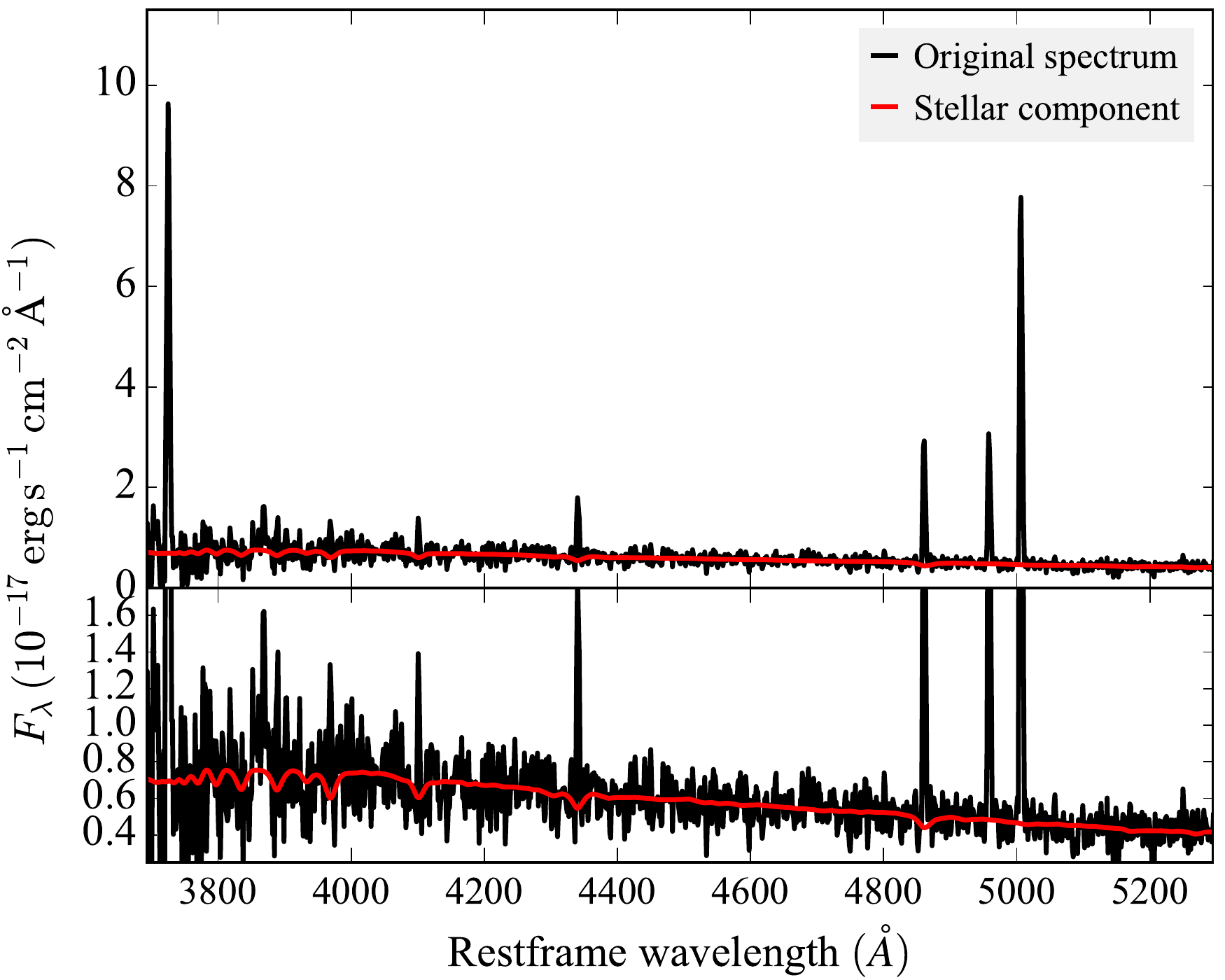}
\caption{FORS2 600$B$ spectrum of the explosion site, i.e., the SN region. {Black is the original spectrum and red represents the fitted stellar component. The \textit{upper panel} shows the full flux range, while the \textit{lower panel} is a zoom into the stellar continuum.}}
\label{fig:FORSSN}
\end{figure}

In particular the 600$B$ spectra (Figs.~\ref{fig:FORSWR} and \ref{fig:FORSSN}) were taken under excellent atmospheric conditions, leading to a width of the spectral line spread function of FWHM=0\farc{6} at 5000~\AA~ as evidenced by the trace of a bright stellar source serendipitously on the slit at a distance of 22\farc{0} to the WR region (seen at $\mathrm{RA(J2000)=19^{h}35^{m}02^{s}.00}$, $\mathrm{Decl(J2000)} = -52^{\circ}50'21\farc{1}$ in Fig.~\ref{fig:Host}). The 600$RI$ data have somewhat worse spatial resolution (FWHM=1\farc{0} at 7000~\AA). This mismatch leads to the complication that the two different spectra are convolved with very different spatial scales. Because of the small angular size of the WR region core ($\lesssim 0\farc{1}$ from HST imaging) and the gradient in physical properties, ratios of lines observed in the two different setups (e.g., for a determination of the dust reddening) are thus clearly nontrivial to interpret.

Table~\ref{tab:fluxes} contains line fluxes and equivalent widths from our analysis of the FORS2 spectra that are sometimes significantly different from the original \citet{2006A&A...454..103H} values. Their actual errors on the line fluxes remain unfortunately unclear, but a substantial uncertainty of at least 25\% must be present; this value was estimated from the obvious discrepancy between their line-flux ratios of \oiii($\lambda$5007)/\oiii($\lambda$4959) and \nii($\lambda$6584)/\nii($\lambda$6548) to the theoretical value of 2.98 set by transition probabilities and observed in high-quality SDSS spectra \citep[e.g.,][]{2000MNRAS.312..813S, 2006agna.book.....O, 2016MNRAS.459.3475W}.

\begin{table*}[!ht]
\caption{Emission-line fluxes and EW measurements from archival FORS2 long-slit spectra for the GRB~980425 host galaxy \label{tab:fluxes}}
\centering
\begin{tabular}{ccccc}
\hline
\hline\noalign{\smallskip}
 &  \multicolumn{2}{c}{SN region$^{a}$} &  \multicolumn{2}{c}{WR region$^{b}$}  \\
 &  Flux$^{c}$ & EW$_{\mathrm{rest}}$ (\AA)  & Flux$^{c}$ & EW$_{\mathrm{rest}}$ (\AA) \\

\hline\noalign{\smallskip}

\oii($\lambda$3727)  & $47\pm2$ & $48\pm3$ & $41\pm3$ & $250\pm5$ \\
\neiii($\lambda$3968)& $4.3\pm0.6$ & $3.1\pm0.9$ & $8.2\pm0.2$ & $52\pm2$ \\
H$\delta$            & $2.6\pm0.5$ & $3.6\pm0.8$ & $7.7\pm0.2$ & $51\pm2$ \\
H$\gamma$            & $5.6\pm0.5$ & $9.0\pm1.2$ & $14.7\pm0.3$ & $57\pm3$ \\
\oiii($\lambda$4363) & $0.6\pm0.2$ & $2.2\pm0.8$ & $1.54\pm0.05$ & $10\pm1$ \\
H$\beta$             & $12.4\pm1.1$ & $24\pm2$ & $34.6\pm1.4$ & $313\pm3$ \\
\oiii($\lambda$4959) & $12.9\pm1.2$ & $24\pm2$ & $68\pm2$ & $650\pm5$ \\
\oiii($\lambda$5007) & $40\pm2$ & $75\pm4$ & $205\pm4$ & $1980\pm20$ \\
\hline
\hline
\nii($\lambda$6548)$^{b}$  & $1.3\pm0.2$ & $5.0\pm1.0$ & $1.57\pm0.03$ & $25\pm1$ \\
H$\alpha$            & $38.5\pm1.6$ & $115\pm6$ & $77\pm2$ & $1220\pm20$ \\
\nii($\lambda$6584)  & $4.9 \pm 0.3$ & $15.9\pm1.8$ & $4.7\pm0.2$ & $79\pm3$ \\
\sii($\lambda$6717)  & $7.8 \pm 0.3$ & $29\pm2$ & $4.4\pm0.2$ & $87\pm3$ \\
\sii($\lambda$6731)  & $5.9 \pm 0.3$ & $20\pm2$ & $3.4\pm0.3$ & $70\pm3$ \\
\hline\noalign{\smallskip}
\end{tabular}

\tablefoot{
\tablefoottext{a}{To increase the S/N ratio, we include the adjacent two pixels in the extraction for the SN region.}
\tablefoottext{b}{Derived from the spectrum at the peak of the emission of the WR region.}
\tablefoottext{c}{Fluxes are given in $10^{-16}~\mathrm{erg}~\mathrm{s}^{1}~\mathrm{cm}^{-2}$.}
\tablefoottext{d}{The double horizontal line separates the nebular lines that were taken in the two different FORS2 observational setups. As a result of the different width of the line spread function and thus angular scales probed in both setups, the emission lines below and above the horizontal line cannot easily be used together to infer physical properties.}
} 
\end{table*}

From the \hb/\hg~ and \hb/\hd~ ratio, we measured $E_{B-V}=0.22\pm0.03$~mag for the WR and $E_{B-V}=0.08_{-0.08}^{+0.23}$~mag for the SN region, which are both perfectly consistent with the MUSE data\footnote{We would measure significant dust reddening if we did not correct our \hb, \hg, or \hd~fluxes for stellar Balmer absorption.}. Also other properties are very similar to our IFU-based values as given in Table~\ref{tab:prop}.

Finally, we exploit the bluer response of the FORS2 600$B$ grism to obtain an age of the \hii~region through a fit using composite stellar templates in \texttt{starlight} in a similar manner as in the main text. The stellar population in the WR region is extremely young and the \citet{2003MNRAS.344.1000B} templates with an ages between 1~Myr and 3~Myr dominate the best fit by contributing $\sim$60\% to 90\% to the total observed starlight in various fits using different spectral templates and reddening laws. This age is consistent with the extremely high EW of \oiii~ and \ha~ (Table~\ref{tab:fluxes}) and the constraints derived in the main text.

The SN region has prominent stellar components with ages of 5 Myr and 40 Myr each contributing around 30\% to the best-fit composite template; the younger of these components is again consistent with the age estimate from the \ha~EW in the main text.

\section{Dependence of O3N2-based oxygen abundance on the ionization parameter}
\label{sec:abundancevsion}

\begin{figure}
\begin{subfigure}{.49\textwidth}
  \includegraphics[width=0.9\linewidth]{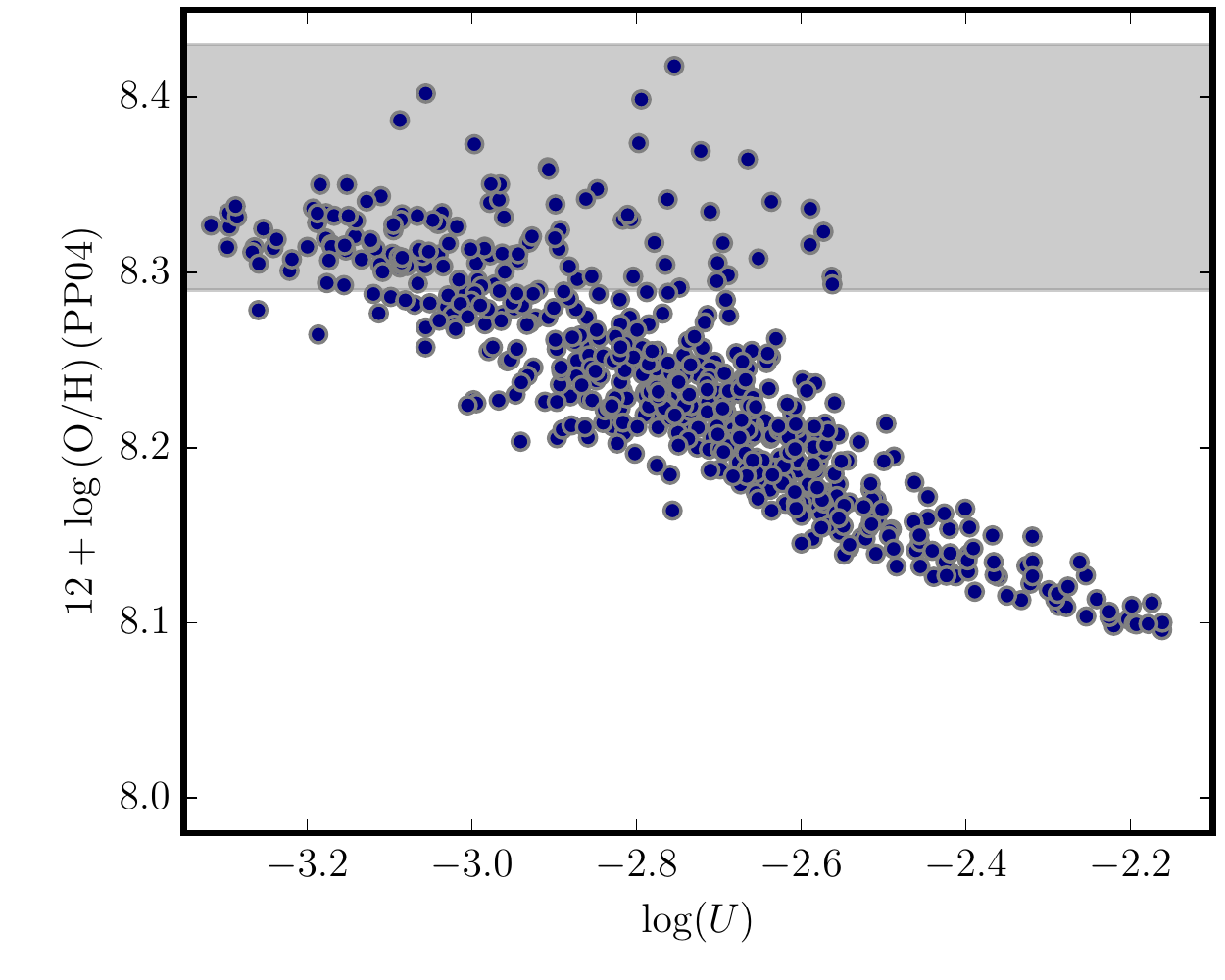}
\end{subfigure}
\begin{subfigure}{.49\textwidth}
  \includegraphics[width=0.9\linewidth]{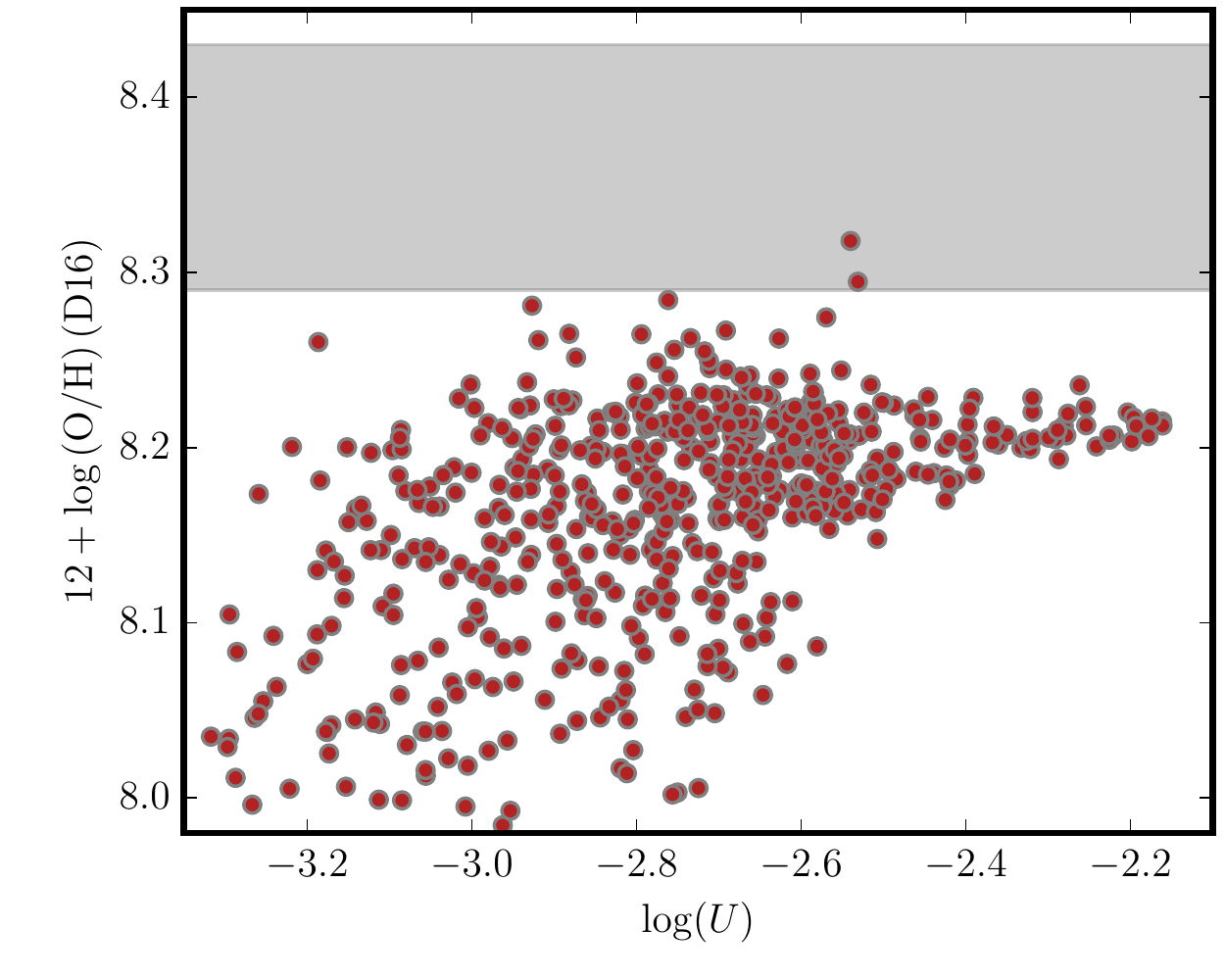}
\end{subfigure}\caption{Dependence of the inferred oxygen abundance in two strong-line diagnostic ratios (\textit{upper panel}: O3N2 from \citealt{2004MNRAS.348L..59P}, \textit{lower panel}: D16) on the ionization parameter $U$ (from \siii/\sii) for the WR region. Each data point corresponds to a single spaxel and the grey region indicates the constraints from the temperature-sensitive \oiii($\lambda$4363) emission line.}
\label{fig:UO3}
\end{figure}

To better illustrate how a changing ionization parameter $U$ affects the metallicity measurement in the different strong-line diagnostics, we plot $U$ (defined as ionizing photons per hydrogen atoms) versus $\oh$~ in the \citet{2004MNRAS.348L..59P} or \citet{2016Ap&SS.361...61D} scale of the brightest \hii\  region in ESO184-G82 in Fig.~\ref{fig:UO3}. Here, we use the \sii/\siii~ratio (Fig.~\ref{fig:s3s2}) to calculate $U$ via photoionization models \citep{2011MNRAS.415.3616D}. Different parameterizations in $U$ \citep[e.g.,][]{2016A&A...594A..37M} do not change Fig.~\ref{fig:UO3} significantly. The \sii/\siii~ratio has the advantage that it is nearly insensitive to metallicity, so we should not measure a strong correlation between both quantities in accurate metallicity diagnostics. It is again clear, however, that the O3N2 metallicity scale only reproduces the oxygen abundance at lower ionization parameters and systematically underpredicts it at higher $U$. In contrast, the \citet{2016Ap&SS.361...61D} diagnostic is mostly independent of ionization parameter, but seems offset by an average $\sim$0.15~dex toward lower $\oh$~ (Fig.~\ref{fig:UO3}, lower panel).

\end{appendix}
\end{document}